\documentclass[10pt]{article}
\usepackage{amsmath,amsfonts,xypic}
\usepackage{amssymb}
\usepackage{amscd}
\usepackage{ulem}  
\usepackage{cancel}
\usepackage[all]{xy}
\usepackage{boxedminipage}
\usepackage[OT2,OT1]{fontenc}
\newcommand\cyr{%
\renewcommand\rmdefault{wncyr}%
\renewcommand\sfdefault{wncyss}%
\renewcommand\encodingdefault{OT2}%
\normalfont
\selectfont}
\DeclareTextFontCommand{\textcyr}{\cyr}

\usepackage[colorlinks=true,urlcolor=blue]{hyperref}
\hypersetup{                
backref = true,             
pagebackref = true,                 
hyperindex = true,          
colorlinks = true,          
breaklinks = true,          
urlcolor = blue,            
linkcolor = blue,           
bookmarks = true,           
bookmarksopen = true,       
citecolor=red,
}
\advance \textheight by \topskip
\makeatletter
\@addtoreset{equation}{section}
 
\makeatother

\newcommand{\nn}{\nonumber}
\newcommand{\beq}{\begin{equation}}
\newcommand{\eeq}{\end{equation}}
\newcommand{\beqa}{\begin{eqnarray}}
\newcommand{\bpm}{\begin{pmatrix}}
\newcommand{\epm}{\end{pmatrix}}
\newcommand{\eeqa}{\end{eqnarray}}
\newcommand{\noi}{\noindent}
\newcommand{\e}{\varepsilon}
\newcommand{\p}{\varphi}
\newcommand{\J}{{\cal J}}
\newcommand{\K}{{\cal K}}

\newcommand{\dd}{\text{d}}
\newcommand{\vt}{\vartheta}

\def\>{\rangle}
\def\<{\langle}

\newcommand{\nc}{\newcommand}
\nc{\rr}{\color{red}}
\nc{\bb}{\color{blue}}
\nc{\gr}{\color{green}}

\begin{document}

\title{{\bf 
Unitary representations of three dimensional  Lie groups
revisited: A short tutorial via harmonic functions 
}}
\author{
{\sf  R. Campoamor-Stursberg}\thanks{e-mail:
rutwig@ucm.es}$\,\,$${}^{a}$ and {\sf  M. Rausch de
Traubenberg}\thanks{e-mail: Michel.Rausch@iphc.cnrs.fr
}$\,\,$${}^{b}$
\\
{\small ${}^{a}${\it I.M.I-U.C.M,
Plaza de Ciencias 3, E-28040 Madrid, Spain}}  \\
{\small ${}^{b}${\it
Universit\'e de Strasbourg, CNRS, IPHC UMR 7178, F-67000 Strasbourg, France
}}\\
{\small {\it  Strasbourg, 67037 Cedex, France}} }
\date{\today}
\maketitle 

\vskip-1.5cm
\begin{abstract}
The representation theory  of three dimensional  real and complex  Lie groups is reviewed from the perspective 
of harmonic functions defined over certain appropriate manifolds. An explicit construction of  all unitary representations is given.  The realisations obtained  are shown to be related with each other by either natural  operations as real forms or 
In\"on\"u-Wigner contractions.
\end{abstract}

\section{Introduction}

\noindent Since its first appearance in the frame of Classical
Mechanics, three dimensional spherical functions have shown to
constitute a valuable and elegant tool at both the mathematical
and physical levels, leading to an extensive and profound theory with multiple ramifications 
in Geometry, Fourier analysis, the theory of special functions and differential equations \cite{Tai,gms,BG,helg, fl,bar,bar2,ggv,lang,WA}. 
In the physical context, spherical functions appear naturally in the construction of bases of
eigenstates in the theory of angular momentum, providing therefore a powerful technique to 
describe accurately the representations of the rotation groups $SO(n)$.  This approach has been central 
to the application of $SO\left( 3\right) $ to Atomic and Nuclear physics, constituting the basis of the theory
of spherical tensor operators and the Racah-Wigner algebraic formalism \cite{FA,JU}. 
The generalization of spherical
functions to higher dimensions and indefinite metric tensors
emerged naturally within the transition of non-relativistic to
relativistic physics, the case of the Lorentz group $SO\left(
1,3\right) $ and the problem of induced/subduced representations with respect to its various subgroups 
being exceptionally important \cite {gms,Nie,Hus,MU1,MU2}. In
this enlarged context, the theory of spherical harmonics establishes an
equivalence between irreducible representations of the rotation groups $%
SO\left( N\right) $ acting on the sphere $\mathbb{S}^{N-1}$ and
eigenfunctions of the spherical Laplacian, while for
pseudo-orthogonal groups $SO\left( p,q\right) $, a similar
formalism has been developed, taking into account their action on
hyperboloids (see e.g. \cite{Str} and references therein).  Beyond
the (pseudo-) orthogonal groups, and usually motivated by specific
physical situations, generalized harmonic or hyper-spherical
functions have been  considered in  the representation theory of
other types of Lie groups, such as the group $SU\left( 3\right) $
in the study of strong interactions \cite{Beg,Nel}, the
quantum mechanics of
three bodies \cite{Marsh}, or the noncompact Lie group $Sp\left( 6,\mathbb{R}%
\right) $ in the frame of the translationally invariant shell
model \cite {Ashe}. Further applications of
the method of harmonics in combination with the internal labeling problem can be found
{\it e.g.} in the nuclear collective model \cite{GR} or the construction of coherent states 
on simple Lie groups \cite{Ino,PER,MA}. It may be mentioned that spherical harmonics,
in combination with Clifford analysis, has recently been extended
to integration problems in superspace \cite{Bie}.

\bigskip
\noindent In this work,  we review the main features of the representation theory of three dimensional semisimple 
Lie groups, emphasising on a complementary/alternative analysis of the representations  from the point of view
of harmonic functions. Following the standard approach, we consider the complex Lie group $SL(2,\mathbb C)$ and its real compact and 
split forms
$SU(2)$ and $SL(2,\mathbb R)$, respectively, revisiting the realisation 
of these groups acting on themselves. These realisations are further identified with what is sometimes called the parameter space of the Lie group, leading in particular to the following identifications:
\beqa
\label{eq:mani}
SL(2,\mathbb C) &\cong& \mathbb S^3_\mathbb C = \Bigg\{z_1,z_2,z_3,z_4 \in \mathbb C,\  z_1^2+z_2^2+z_3^2 +z_4^2 =1 \Bigg\}\ ,\nn\\
SU(2) &\cong& \mathbb S^3  = \Bigg\{x_1,x_2,x_3,x_4 \in \mathbb R,\  x_1^2+x_2^2+x_3^2 +x_4^2 =1 \Bigg\}\ ,\\
SL(2,\mathbb R) &\cong&  \mathbb H_{2,2}= \Bigg\{x_1,x_2,x_3,x_4 \in \mathbb R,\  x_1^2+x_2^2-x_3^2 -x_4^2 =1 \Bigg\}\nn \ .
\eeqa
 Considering a complex vector space (respectively a manifold) $E$ (resp. $\cal M$)  and a real vector space  
(resp. a real manifold) $F$
(resp. $\cal N$), we  call  $F$  (resp. ${\cal N}$)  a
real form of the vector space $E$ (resp. the manifold $\cal M$) 
 whenever the isomorphisms $E\cong  F\otimes_\mathbb R \mathbb C$ (resp.  ${\cal M} \cong {\cal N} \otimes_\mathbb R  \mathbb C$ locally)  hold.
In particular, when the manifolds  $\cal M$ and $\cal N$ have a structure of complex or real Lie group this
definition coincides with the usual definition of real forms of Lie groups \cite{wyb,GIL}. 
With this terminology,  we have $\mathbb S^3_\mathbb C \cong \mathbb S^3  \otimes_\mathbb R \mathbb C 
\cong \mathbb H_{2,2} \otimes_\mathbb R \mathbb C $ or,  alternatively at 
the Lie group level,
$SL(2,\mathbb C) \cong SU(2) \otimes_\mathbb R \mathbb C \cong SL(2,\mathbb R) \otimes_\mathbb R \mathbb C$. 
By abuse of language we  shall  call $\mathbb S^3_\mathbb C$
the ``complex unit three-sphere''. 
 Endowing the manifolds \eqref{eq:mani} with a system of coordinates, the parameterisation turns out to be
bijective if appropriate regions are removed (circles ${\mathbb S}^1$ or cylinders ${\cal C}_2$) leading to
the following bijective parameterisations
\beqa
\begin{array}{ccc}
\mathbb S^3_\mathbb C \setminus\Big({\cal C}_2 \times {\cal C}_2\Big) &\text{for}&SL(2,\mathbb C) \ , \\
\mathbb S^3\setminus\Big(\mathbb S^1 \times \mathbb S^1\Big) &\text{for}&SU(2) \ , \\
 \mathbb H_{2,2} \setminus \mathbb S^1&\text{for}&SL(2,\mathbb C) \ .\nn
\end{array}
\eeqa
 Using these  parameterisations, convenient differential realisation of
the Lie algebras $\mathfrak{sl}(2,\mathbb C), \mathfrak{su}(2)$ and $\mathfrak{sl}(2,\mathbb R)$ can be obtained.
The  main feature of these parameterisation lie in their interpretation  in the sense of real forms of vector spaces
(manifolds) given above. Indeed, it  turns out  that the differential realisations defined 
 for the Lie algebras $\mathfrak{su}(2)$ or $\mathfrak{sl}(2,\mathbb R)$ are obtained from the
differential realisation defined for $\mathfrak{sl}(2,\mathbb C)$, considering
 appropriate real forms. 
Next, with these  explicit differential realisation  we reconsider unitary irreducible  representations
of these Lie groups, within a unified formulation using the corresponding
manifolds \eqref{eq:mani}.
The case of $SL(2,\mathbb R)$, as known \cite{lang}, is somewhat peculiar,   as 
 $SL(2,\mathbb R)$ admits some ($p-$sheeted, universal) covering groups.
 This differential realisation of $SL(2,\mathbb R)$ is seen to extend to 
  a suitable covering of  $\mathbb H_{2,2} \setminus \mathbb S^1$ 
in such a way that it acts naturally 
on unitary representations of the corresponding covering of  
 $SL(2,\mathbb R)$. In this context, the preceding observations show that the unitary representations of $\mathfrak{sl}(2,\mathbb C)$ can
be conveniently  adapted to cover the various real forms in order to derive a description of representations in terms of Gel'fand formul\ae\
, applied to the appropriate manifolds.  

The above-mentioned formalism can also be reformulated to deal with the unitary 
representations of the non-compact, non-semisimple
Euclidean group $E_2$, by means of appropriate contractions of realisations.
It is well known 
that the Lie group $E_2$ can be obtained by some contractions of $SU(1,1)$ \cite{IW,WE}. This relation remains also valid
at the level of the parameter space. 
The parameter space of $E_2$ is  $\mathbb R^2 \times [0,2\pi[$ which
is related to
$[0,2\pi[\times [0,2\pi[ \times \mathbb R_+$ and allows a parameterisation of the 
 cone ${\cal C}_{2,2}$ of equation
\beqa
x_1^2 +x_2^2 -x_3^2 - x_4^2 = 0 \ . \nn
\eeqa
The parameterisation becomes bijective on ${\cal C}_{2,2} \setminus\{0\}$, {\it i.e.}, on the cone with one
point removed.
 Since ${\cal C}_{2,2} \setminus\{0\} $ can be seen as a singular limit of ${\mathbb H}_{2,2}\setminus \mathbb S^1$,
 all unitary representations of the Euclidean group $E_2 $ are obtained by a contraction procedure of  $SU(1,1)$ representations as functions on 
${\cal C}_{2,2}\setminus \{0\}$.  Such a process to obtain Lie groups with a semi-direct structure from semisimple Lie groups
is known as confluence (see {\it e.g.} \cite{ww}).

\medskip
\noindent 
We recall that in \cite{nor}, realisations  of the Lie groups $SO(3), SO(1,2)$ and $E_2$  have been obtained in unified 
manner for the description of the Landau quantum systems, 
albeit these realisations differ slightly from the construction considered here, which mainly reviews and exploits the geometric features of representations.  The main purpose of this reformulation is to point out the correspondence with the theory of
harmonic functions \cite{ggv,Tai,WA}, in a practical direct way, deprived of the usual heavy machinery of mathematical monographs, and hence suitable for application in various physical problems that need of the representation theory, but not necessarily using the latter to its last formal consequences. 

\medskip
\noindent  The review of the representation theory along these lines can be graphically summarised in the following diagram
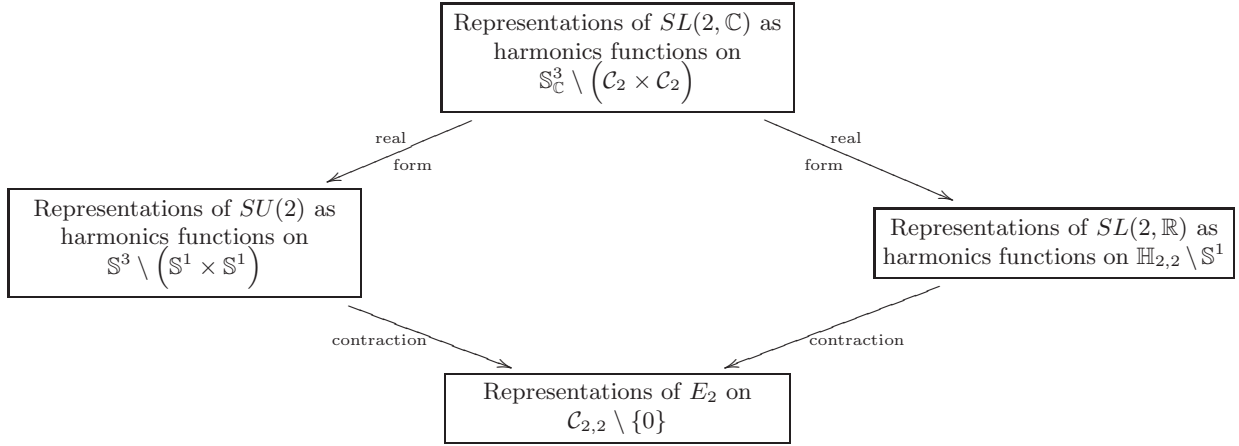
\small{
\beqa
\xymatrix{
& 
\begin{boxedminipage}{4.7cm}
\begin{center}
Representations of  $SL(2,\mathbb C)$
as  harmonics functions on $\mathbb S^3_\mathbb C \setminus\Big({\cal C}_2\times {\cal C}_2\Big)$
\end{center}
\end{boxedminipage}\ar[dl]_{\text{real}}^{\text{form}}\ar[dr]^{\text{real}}_{\text{form}}\\
\begin{boxedminipage}{4.7cm}
\begin{center}
Representations of  $SU(2)$
as harmonics functions on $\mathbb S^3 \setminus \Big(\mathbb S^1\times \mathbb S^1\Big)$
\end{center}
\end{boxedminipage}\ar[dr]_{\text{contraction}}&&
\begin{boxedminipage}{4.8cm}
\begin{center}
\small Representations of  $SL(2,\mathbb R)$
as harmonics functions on $\mathbb H_{2,2}\setminus \mathbb S^1$
\end{center}
\end{boxedminipage}\ar[dl]^{\text{contraction}}\\
&\begin{boxedminipage}{4.6cm}
\begin{center}
Representations of  $E_2$
on  ${\cal C}_{2,2}\setminus\{0\}$
\end{center}
\end{boxedminipage}
}
\nn
\eeqa
}


\section{The $\mathfrak{sl}(2,\mathbb C)$ algebra}

\noindent We briefly review the main facts concerning the representation theory of the Lie algebra $\mathfrak{sl}(2,\mathbb C)$ and
the theory of harmonic functions, and fix some of the notations that will be used in later paragraphs. For details on the construction the reader is referred 
to the main references \cite{gms,ggv,nai2} (see \cite{nai-eng} for an English translation of \cite{nai2}).
 Other alternative constructions of the representations 
can be found in \cite{VK1,VK2,VK3}.

\subsection{Linear   representations of $\mathfrak{sl}(2,\mathbb C)$}
The real $\mathfrak{sl}(2,\mathbb C)$ algebra is the     six-dimensional Lie algebra generated by
\footnote{The Lie group $SL(2,\mathbb C)$ is the universal covering group of the Lorentz group $SO(1,3)$
generated by $J_{\mu \nu} = -J_{\nu \mu}, 0 \le \mu, \nu \le 3$  and satisfying
\beqa
[J_{\mu \nu}, J_{\rho \sigma}]= -i (\eta_{\mu\rho} J_{\nu\sigma} -\eta_{\nu \rho} J_{\mu \sigma} + \eta_{\mu\sigma} J_{\rho \nu}
-\eta_{\nu\sigma} J_{\rho\mu}) \ , \nn
\eeqa
with  $\eta_{\mu \nu} = \text{diag}(1,-1,-1,-1)$. The relationship with the generators \eqref{eq:geneLor}
is given by $J_0=J_{12}, J_+ = J_{23}+i J_{31}, J_- =J_{23}-i J_{31}$ (generators of rotations)
and $K_0=J_{03}, K_+= J_{01}+i J_{02},
K_-= J_{01}-i J_{02}$ (generators of Lorentz boosts).}
$J_\pm,J_0,K_\pm, K_0$ with non-vanishing commutation relations
\beqa
\label{eq:geneLor}
\begin{array}{lll}
\big[J_0,J_+\big]=J_+\ ,&\big[J_0,K_+\big]=K_+\ ,& \big[K_0,K_+\big]=-J_+\ ,\\
&\big[K_0,J_+\big]=K_+\ ,&\\
\big[J_0,J_-\big]=-J_-\ ,&\big[J_0,K_-\big]=-K_-\ ,& \big[K_0,K_-\big]=J_-\ ,\\
&\big[K_0,J_-\big]=-K_-\ ,&\\
\big[J_+,J_-\big]=2J_0\ ,&\big[J_+,K_-\big]=2K_0\ ,& \big[K_+,K_-\big]=-2J_0\ ,\\
&\big[J_-,K_+\big]=2K_0\ .
\end{array}
\eeqa
The two (quadratic) Casimir operators are given by
\beqa
Q_1 &=& J_0^2 + \frac12(J_+ J_- + J_- J_+) - K_0^2 -\frac12(K_+ K_- + K_- K_+) \ , \nn \\
Q_2 &=& J_0 K_0 + \frac12(J_+ K_- + J_- K_+) +  K_0 J_0 + \frac12(K_+ J_- + K_- J_+) \ . \nn
\eeqa

\noi
Introducing the Pauli matrices
\beqa
\sigma_1=\bpm 0&1\\1&0 \epm \ , \ \
\sigma_2=\bpm 0&-i\\i&\phantom{-}0 \epm \ , \ \
\sigma_3=\bpm 1&\phantom{-}0\\0&-1 \epm \ , \nn
\eeqa
$\sigma_\pm = \sigma_1 \pm i \sigma_2$,  $z^1, z^2 \in \mathbb C$ and
\beqa
Z = \bpm z^1&z^2\epm\ , \ \
\bar Z = \bpm \bar z_1 \\ \bar z_2\epm\ , \ \
\partial_Z= \bpm \partial_1\\\partial_2\epm \ , \ \
\partial_{\bar Z}= \bpm \bar \partial^1 & \bar \partial^2\epm \ , \nn
\eeqa
a very convenient and efficient realisation  can be given   by
\beqa
\J_+&=& Z \sigma_+ \partial_Z \ - \partial_{\bar Z}\sigma_+ {\bar Z} = z^1 \partial_2-\bar z_2 \bar \partial^1\ , \nonumber \\
\J_-&=& Z \sigma_- \partial_Z \ - \partial_{\bar Z} \sigma_- {\bar Z} =z^2 \partial_1-\bar z_1 \bar \partial^2 \ ,  \nn\\
\J_0& =&\frac12   Z \sigma_3 \partial_Z-\frac12   \partial_{\bar Z} \sigma_3 {\bar Z} =
\frac12(z^1\partial_1 - z^2 \partial_2)-\frac12(\bar z_1\bar\partial^1 - \bar z_2 \bar\partial^2) \ ,\nn \\
\K_+&=& i(Z\sigma_+ \partial_Z \ + \partial_{\bar Z}\sigma_+ {\bar Z}) \ = i
 \Big(z^1 \partial_2+\bar z_2 \bar \partial^1\big)\ , \nn\\
\K_-&=& i(Z\sigma_- \partial_Z \ + \partial_{\bar Z}\sigma_- {\bar Z}) \ = i
\Big( z^2 \partial_1+\bar z_1 \bar \partial^2\ \Big)\ , \nn\\
\K_0&=&\frac{i}2   Z \sigma_3 \partial_Z+\frac{i}2   \partial_{\bar Z} \sigma_3 {\bar Z} =
\frac{i}2(z^1\partial_1 - z^2 \partial_2)+\frac{i}2(\bar z_1\bar\partial^1 - \bar z_2 \bar\partial^2) \ .
\nonumber
\eeqa
With these notations, representations of $\mathfrak{sl}(2,\mathbb C)$ are given by the set of homogeneous functions
in two complex variables \cite{ggv}
\beqa
{\cal D}_{p,q} =\Bigg\{f  \quad | \quad  \ \  f(\lambda z^1, \lambda z^2,\lambda \bar z_1, \lambda \bar z_2)=
\lambda^p \bar \lambda^q f( z^1, z^2, \bar z_1,  \bar z_2) \Bigg\} \ . \nn
\eeqa
In order to avoid  the  monodromy problem, we have to assume that  the condition $p-q \in \mathbb Z$  holds \cite{ggv}.

\medskip
\noindent
 The  irreducible representations  were originally obtained by Gel'fand  (see e.g. \cite{gms,nai,gt,hc} and references therein),
and are characterised by two numbers $\ell_0, \ell_1$,  whereas the pair $[\ell_0,\ell_1]$ denotes the representation.
 Explicitly, they are given by the set of functions
\cite{st}
\beqa
\label{eq:lor}
\psi_{\ell_0,\ell_1}^{s,m}(Z,\bar Z)&=& A_{s}^{\ell_0,\ell_1} \sqrt{(2s+1)(s+m)!(s-m)!(s+\ell_0)!(s-\ell_0)!}\times\nn\\
&&\times (Z \bar Z)^{\ell_1-s-1}\sum \limits_k
\frac{(z^1)^{m+\ell_0+k} (-\bar z_1)^k (z^2)^{s-m-k} (\bar z_2)^{s-\ell_0-k}}
     {(m+\ell_0+k)! k ! (s-m-k)! (s-\ell_0 -k)!} \ ,
\eeqa
 where
\beqa
A_s^{\ell_0,\ell_1}=\sqrt{\frac{\Gamma(s-\ell_1+1)\Gamma(|\ell_0|+\ell_1+1)}{\Gamma(s+\ell_1+1)\Gamma(|\ell_0|-\ell_1+1)}}=
\sqrt{\frac{(s-\ell_1) \cdots(|\ell_0|+\ell_1+1)}{(s+\ell_1)\cdots(|\ell_0|+\ell_1+1)}} \ . \nn
\eeqa
In the  sum \eqref{eq:lor}, the index $k$ is such that all powers are positive. This in particular  implies the constraint
\beqa
\text{max}(0,-\ell_0-m)\le k \le \text{min}(s-\ell_0,s-m) \ . \nn
\eeqa
Observe that the functions $\psi_{\ell_0,\ell_1}^{s,m}$ belong to the space ${\cal D}_{\ell_0+\ell_1-1,-\ell_0+\ell_1-1}$, thus we must
have $2\ell_0\in \mathbb Z$,  and hence  $s= |\ell_0|, |\ell_0| +1,\cdots$ and $-s\le m \le s$.
 The following isomorphism of representations holds \cite{ggv}
\beqa
[\ell_0,\ell_1] \cong [ -\ell_0,-\ell_1]  \ . \nn
\eeqa
Since $\ell_0 \in 2\mathbb Z$, we assume now that $\ell_0\ge0$. A representation of $\mathfrak{sl}(2,\mathbb C)$
is then characterised by a positive half-integer number $\ell_0$ and a complex number $\ell_1$.

The action of the $\mathfrak{sl}(2,\mathbb C)$ generators \eqref{eq:geneLor} gives \cite{st}
{\small
\beqa
\J_+ \psi_{\ell_0,\ell_1}^{s,m}(Z,\bar Z) &=&
\sqrt{(s-m)(s+m+1)} \psi_{\ell_0,\ell_1}^{s,m+1}(Z,\bar Z) \nn \ , \\
\J_- \psi_{\ell_0,\ell_1}^{s,m}(Z,\bar Z) &=&
\sqrt{(s+m)(s-m+1)} \psi_{\ell_0,\ell_1}^{s,m-1}(Z,\bar Z) \nn \ , \\
\J_0 \psi_{\ell_0,\ell_1}^{s,m}(Z,\bar Z) &=& m \psi_{\ell_0,\ell_1}^{s,m}(Z,\bar Z) \ , \\
\K_+ \psi_{\ell_0,\ell_1}^{s,m}(Z,\bar Z) &=&\phantom{+}C_s \sqrt{(s-m)(s-m-1)}\psi_{\ell_0,\ell_1}^{s-1,m+1}(Z,\bar Z)
-i\frac{\ell_0 \ell_1}{s(s+1)}\sqrt{(s-m)(s+m+1)}\psi_{\ell_0,\ell_1}^{s,m+1}(Z,\bar Z) \nn\\
&&+C_{s+1} \sqrt{(s+m+1)(s+m+2)} \psi_{\ell_0,\ell_1}^{s+1,m+1}(Z,\bar Z) \ , \nn\\
K_- \psi_{\ell_0,\ell_1}^{s,m}(Z,\bar Z) &=&-C_s \sqrt{(s+m)(s+m-1)}\psi_{\ell_0,\ell_1}^{s-1,m-1}(Z,\bar Z)
-i\frac{\ell_0 \ell_1}{s(s+1)}\sqrt{(s+m)(s-m+1)}\psi_{\ell_0,\ell_1}^{s,m-1}(Z,\bar Z) \nn\\
&&-C_{s+1} \sqrt{(s-m+1)(s-m+2)} \psi_{\ell_0,\ell_1}^{s+1,m-1}(Z,\bar Z) \ , \nn\\
\K_0 \psi_{\ell_0,\ell_1}^{s,m}(Z,\bar Z) &=& C_s\sqrt{(s-m)(s+m)}\psi_{\ell_0,\ell_1}^{s-1,m}(Z,\bar Z)
-i\frac{\ell_0 \ell_1}{s(s+1)} \psi_{\ell_0,\ell_1}^{s,m}(Z,\bar Z)\nn \\
&&-C_{s+1}\sqrt{(s+m+1)(s-m+1)} \psi_{\ell_0,\ell_1}^{s+1,m}(Z,\bar Z) \ , \nn
\eeqa
}
 where
\beqa
C_s=\frac{i}{s}\sqrt{\frac{(s^2-\ell_0^2)(s^2-\ell_1^2)}{4s^2-1}} \ . \nn
\eeqa
The Casimir operators are given  respectively by
\beqa
\label{eq:Cas-sl2C}
Q_1 &=& \ell_0^2 + \ell_1^2 -1 \ ,\nn \\
Q_2 &=& -2i \ell_0 \ell_1 \ .
\eeqa
We observe that for a given $s=\ell_0,\cdots,$ the functions $\psi_{\ell_0,\ell_1}^{s,-s},\cdots,\psi_{\ell_0,\ell_1}^{s,s}$ span
the spin$-s$ representation of the   subalgebra $\mathfrak{su}(2) \subset \mathfrak{sl}(2,\mathbb C)$.  
The representation $[\ell_0,\ell_1]$ is  generally infinite dimensional. However, if both
$\ell_0$ and $\ell_1$ are simultaneously   half-integers  and $\ell_1\ge \ell_0 +1$, the representation  is finite
dimensional and its spin content is given by $s= \ell_0,\ell_0+1,\cdots, \ell_1-1$. In this case we
observe that $C_{\ell_0}=C_{\ell_1}=0$, or  $p=\ell_0+\ell_1-1,q=-\ell_0+\ell_1-1$ are both integers and positive.
Moreover, the power of $Z \bar Z$ in \eqref{eq:lor} ranges from  $\ell_1-1$ to $0$ and thus
is always positive.
To make contact with more familiar notations, the finite dimensional  representations  can be rewritten as
\beqa
{\cal D}_{p,q}= \Big\{ (z^1)^{p-m} (z^2)^m (\bar z_2)^{q-n} (\bar z_1)^n, 0\le m \le p, 0 \le n\le q \Big\} \ , \nn
\eeqa
with $p=\ell_0+\ell_1 -1, q=\ell_1-\ell_0-1$.
In particular, ${\cal D}_{1,0}\cong[1/2,3/2]$ and ${\cal D}_{0,1}\cong[-1/2,3/2]\cong[1/2,-3/2]$ correspond respectively to left- or right-handed spinors.

\medskip
\noindent  We recall that  the representation is unitary \cite{gms,nai,gt,hc} whenever one of the following conditions holds:
\begin{enumerate}
\item $\ell_0 \in\frac12 \mathbb N$ and $\ell_1 = i \sigma, \sigma\in \mathbb R$ (principal series);
\item $\ell_0 =0$ and $0< \ell_1\le 1$ (complementary series).
\end{enumerate}

\noindent For  the principal series, the Hilbert space is defined as follows:  
Replacing $z_1 \to z$ and $z_2 \to 1$ in \eqref{eq:lor} , the Hilbert space turns out to coincide with the space of square integrable functions $L^2(\mathbb C)$, 
where
\beqa
(f,g) = \frac1{\pi} \int \limits_{\mathbb R^2} \bar f(\bar z) g(z) dx dy \ ,\nn
\eeqa
with $z = x+iy$.  The  functions $\Psi^{s,m}_{\ell_0,\ell_1}$ are orthogonal  with respect
to this scalar product \cite{st}. The case of the complementary series   is much more involved, having been analysed in detail in
\cite{ggv,nai2,nai-eng}, where explicit 
constructions can be found.

\subsection{Realisation of the representations of $\mathfrak{sl}(2,\mathbb C)$ on 
 $\mathbb S_\mathbb C^3 \setminus\Big({\cal C}_2\times
{\cal C}_2\Big)$ }\label{sec:S3C}
The $SL(2,\mathbb C)-$Lie group is the set of two-by-two unimodular complex
matrices of determinant one
\beqa
U=\bpm \alpha & \beta \\ \gamma& \delta\epm \ , \ \ \alpha,\beta,\gamma, \delta\in
\mathbb C \ , \ \     \alpha\delta-\beta \gamma=1 \ . \nn
\eeqa
{\\ Setting} $\alpha=z_1 - i z_2, \delta = z_1 + i z_2, \beta=z_4 - i z_3, \gamma=-z_4 -i z_3$,
 it is straightforward to verify that  $SL(2,\mathbb C)$ can be identified with the complex unit sphere $\mathbb S^2_\mathbb C$.  

In the previous section we have briefly reviewed the Gel'fand construction of  unitary  representations of
$\mathfrak{sl}(2,\mathbb C)$ in terms of appropriate homogeneous functions on ${\mathbb C}^2$. 
 The next natural step is to extend the Gel'fand formul\ae \  to the space $\mathbb S^2_\mathbb C$. 
 This in particular leads to a unified description of  the $\mathfrak{sl}(2,\mathbb C)$ representations
 by means of functions on ${\mathbb S}_\mathbb C^3$. 
This turns out to be equivalent to finding a differential realisation of $\mathfrak{sl}(2,\mathbb C)$  acting on
$SL(2,\mathbb C)$ by a left action and determining left-invariant vector fields.

\noi
The complex unit three-sphere  itself can obtained by  complexification
(in the sense given in the introduction) of the  ordinary three-sphere, which can be parameterised 
by $\vt_0 \in [0,\pi/2],  \p_{\pm 0} \in [0,2\pi[$ (see Section \ref{sec:unisu2}).
 Introducing  the complexification of the previous angles
$\Theta= \vt_0 + i\vt_1\ , \Phi_\pm = \p_{\pm 0}+i \p_{\pm 1}$ (with $\vt_1,\p_{\pm 1} \in \mathbb R$)
and  defining the four complex numbers
\beqa
z_c= \cos \Theta\ , \ \ z_s= \sin \Theta\ \ , \ \
\zeta_{\pm} = e^{i \Phi_\pm} \ , \nn
\eeqa
 the complex unit  three-sphere ${\mathbb S}^3_\mathbb C  \subset \mathbb C^4$
can be  parameterised by
\beqa
\label{eq:s3c}
z_+ &=& z_c  \zeta_+ = \cos(\Theta) e^{i \Phi_+} \ , \nn \\
z_- &=& z_s \zeta_- = \sin(\Theta) e^{i \Phi_-} \ , \nn \\
z'_+ &=& \frac{\bar z_c}{ \bar \zeta_+} = \cos(\bar \Theta) e^{i \bar \Phi_+} \ ,  \\
z'_- &=& \frac{\bar z_s}{ \bar \zeta_-} = \sin(\bar\Theta) e^{i \bar \Phi_-} \ , \nn
\eeqa
leading to  the identity
\beqa
\label{eq:S3C}
z_+ \bar z'_+ + z_- \bar z'_- =1 \ .
\eeqa
Equation \eqref{eq:S3C} implies in particular that  $z_+ \bar z'_+ =\cos^2 \Theta,z_- \bar z'_- =\sin^2 \Theta$ 
with $\Theta \in \mathbb C$.
Starting from $z_+=\bar z'_+ =\cos \Theta$ and $z_- = \bar z'_- = \sin \Theta$, generic points can be obtained 
as follows:
\beqa
z_\pm \to w_\pm z_\pm\ ,\ \  \bar z'_\pm \to \frac {\bar z_\pm}{w_\pm} \  , \nn
\eeqa
with $w_\pm \ne 0$. Thus we can choose $w_\pm = \zeta_\pm$  leading to the parameterisation
   \eqref{eq:s3c}.  
Finally, using the properties  of the  elementary trigonometric functions  it is sufficient to assume
\beqa
(\vt_0, \p_{+0}, \p_{-0}, \vt_1,  \p_{+1}, \p_{-1}) \in [0,\frac{\pi}2]\times [0,2\pi[ \times  [0,2\pi[ \times \mathbb R^3
\ . \nn
\eeqa
Thus, the  parameterisation above  covers $\mathbb S^3_\mathbb C$. If we now remove the two cylinders ${\cal C}_2$ from $\mathbb S^3_\mathbb C$
defined by
\beqa
\Theta&=&0\ , \ \ (\p_{+0}, \p_{+1}) \in [0,2\pi[ \times \mathbb R \  , \nn \\
\Theta&=&\frac{\pi}2\ , \ \ (\p_{-0} , \p_{-1}) \in [0,2\pi[ \times \mathbb R \  , \nn 
\eeqa
and we denote 
${\cal I}^3_\mathbb C = \Big\{
(\vt_0, \p_{+0}, \p_{-0}, \vt_1,  \p_{+1}, \p_{-1}) \in [0,\frac{\pi}2]\times [0,2\pi[ \times  [0,2\pi[ \times \mathbb R^3
\ , \text{s.t.}\ (\vt_0, \vt_1)\ne (0,0), (\pi/2,0) \Big\}$
 we obtain a bijective map from ${\cal I}^3_\mathbb C$ onto 
$\mathbb S^3_\mathbb R \setminus \Big({\cal C}_2 \times {\cal C}_2\Big)$. 
We observe that the application from  ${\cal I}^3_\mathbb C$ to $\mathbb S^3_\mathbb R \setminus \Big({\cal C}_2 \times {\cal C}_2\Big)$ is continuous, although the reciprocal application is obviously not continuous, for which reason the considered map
does not constitute a homeomorphism.

A direct computation shows that the differential operators defined on ${\mathbb S}_\mathbb C^3 \setminus 
\Big({\cal C}_2 \times {\cal C}_2\Big) $
\beqa
\label{eq:liesl2c}
L_+&=&\frac14 e^{i(-\p_{-0}-i \p_{-1} + \p_{+0} + i \p_{+1})}\Big(
\tan(\vt_0 + i \vt_1) \big(-i \partial_{\p_{+0}} -\partial_{\p_{+1}}\big) +\nn\\
&&
\partial_{\vt_0} - i \partial_{\vt_1} + \cot(\vt_0 + i \vt_1) \big(-i \partial_{\p_{-0}} -\partial_{\p_{-1}}\big)\Big) \nn \\
&=& \frac12 e^{i(\Phi_+ - \Phi_-)} \Big(-i \tan \Theta \partial_{\Phi_+}  + \partial_\Theta-i \cot\Theta \partial_{\Phi_-}\Big) \ , \nn\\
L_-&=&\frac14 e^{i(\p_{-0}+i \p_{-1} - \p_{+0} - i \p_{+1})}\Big(
\tan(\vt_0 + i \vt_1) \big(-i \partial_{\p_{+0}} -\partial_{\p_{+1}}\big) +\\
&&
-\partial_{\vt_0} + i \partial_{\vt_1} + \cot(\vt_0 + i \vt_1) \big(-i \partial_{\p_{-0}} -\partial_{\p_{-1}}\big)\Big) \nn \\
&=& \frac12 e^{i(\Phi_- -\Phi_+)} \Big(-i \tan \Theta \partial_{\Phi_+}  - \partial_\Theta-i \cot\Theta \partial_{\Phi_-}\Big) \ ,
 \nn\\
L_0&=&-\frac{i}4\Big(\partial_{\phi_{+0}} -i\partial_{\p_{+1}} -\partial_{\p_{-0}} +i\partial_{\p_{-1}}\Big) =
-\frac{i}2 \Big(\partial_{\Phi_+} - \partial_{\Phi_-} \Big) \ , \nn
\eeqa
and
\beqa
\label{eq:liesl2c-2}
\bar L_+&=&\frac14 e^{i(-\p_{-0}+i \p_{-1} + \p_{+0} - i \p_{+1})}\Big(
\tan(\vt_0 - i \vt_1) \big(-i \partial_{\p_{+0}} +\partial_{\p_{+1}}\big) +\nn\\
&&
\partial_{\vt_0} + i \partial_{\vt_1} + \cot(\vt_0 - i \vt_1) \big(-i \partial_{\p_{-0}} +\partial_{\p_{-1}}\big)\Big) \nn \\
&=& \frac12 e^{i(\bar\Phi_+ - \bar\Phi_-)} \Big(-i \tan \bar\Theta \partial_{\bar\Phi_+}  + \partial_{\bar\Theta}-
i \cot\bar\Theta \partial_{\bar\Phi_-}\Big) \ , \nn\\
\bar L_-&=&\frac14 e^{i(\p_{-0}-i \p_{-1} - \p_{+0} + i \p_{+1})}\Big(
\tan(\vt_0 - i \vt_1) \big(-i \partial_{\p_{+0}} +\partial_{\p_{+1}}\big) +\\
&&
-\partial_{\vt_0} - i \partial_{\vt_1} + \cot(\vt_0 - i \vt_1) \big(-i \partial_{\p_{-0}} +\partial_{\p_{-1}}\big)\Big) \nn \\
&=& \frac12 e^{i(\bar\Phi_- -\bar\Phi_+)} \Big(-i \tan \bar\Theta \partial_{\bar\Phi_+}
- \partial_{\bar\Theta}-i \cot\bar\Theta \partial_{\bar\Phi_-}\Big) \ ,
 \nn\\
\bar L_0&=&-\frac{i}4\Big(\partial_{\p_{+0}} +i\partial_{\p_{+1}} -\partial_{\p_{-0}} -i\partial_{\p_{-1}}\Big) =
-\frac{i}2 \Big(\partial_{\bar\Phi_+} - \partial_{\bar \Phi_-} \Big) \ , \nn
\eeqa
satisfy the $\mathfrak{sl}(2,\mathbb C)$  commutation relations
\beqa
&\big[L_0,L_\pm\big] = \pm L_{\pm} \ , \ \ \big[L_+,L_-\big] = 2 L_{0} \ , \nn \\
&\big[\bar L_0,\bar L_\pm\big] = \pm \bar L_{\pm} \ , \ \ \big[\bar L_+,\bar L_-\big] = 2 \bar L_{0} \ , \nn \\
&[L_a,\bar L_b\big]=0 \ . \nn
\eeqa
 In order to make contact with the notations considered previously, we observe that
\beqa
J_\pm = L_\pm + \bar L_\pm\ , J_0= L_0 + \bar L_0 \ , \ \
K_\pm = -i(L_\pm - \bar L_\pm)\ , K_0= -i(L_0 - \bar L_0) \ . \nn
\eeqa

\medskip
\noindent
Within this realisation, it turns out that the  spinor representations are given by
\beqa
{\cal D}_{1,0} &=& \Big\{z_-,z_+\Big\} \cong \Big\{\bar z'_+, \bar z'_-\Big\} \ ,  \nn \\
{\cal D}_{0,1} &=& \Big\{\bar z_+,\bar z_1\Big\} \cong \Big\{ z'_-,  z'_+\Big\} \ .  \nn
\eeqa
 In particular, the action is given by
\beqa
\begin{array}{llll}
J_+ z_+=0  \ , & J_+z_-=z_+ \ , &      J_+ \bar z'_+=-\bar z_-' \ , &J_+ \bar z'_-=0 \ , \\
J_- z_+=z_-\ , &  J_- z_-=0 \ , &     J_-\bar z'_+=0\ , &J_- \bar z'_+=-\bar z'_+\ , \\
J_0 z_+ = \frac12 z_+\ , &        J_0  z_- = -\frac12 z_-\ , &
J_0\bar z'_+=-\frac12 \bar z'_+\ , &J_0 \bar z'_- = \frac12 \bar z'_- \ , 
\end{array}
\eeqa
for the left-handed spinors,  whereas for the right-handed spinors we obtain
\beqa
\begin{array}{llll}
\bar J_+ z'_+=0  \ ,    &\bar J_+ z'_-= z'_+\ , &      \bar J_+  \bar z_+=-\bar z_-\ , &\bar J_+ \bar z_-=0\ , \\
\bar J_-  z'_+=z'_-\ , &\bar J_- z'_-=0    \ ,       & \bar J_- \bar z_+=0\ , &\bar J_- \bar z_-=-\bar z_+\ , \\
\bar J_0 z'_+ = \frac12  z'_+\ , &\bar J_0 z'_- = -\frac12  z'_-\ , &
\bar J_0\bar z_+=-\frac12 \bar z_+\ , &\bar J_0 \bar z_- = \frac12 \bar z_- \ .
\end{array}
\eeqa
This explicitly shows that the complex unit
 three-sphere \eqref{eq:S3C} is invariant under the action of the generators of
$\mathfrak{sl}(2,\mathbb C)$,  hence
 we are able to extend safely the unitary representations 
 to  ${\mathbb S}^3_\mathbb C\setminus\Big({\cal C}_2 \times {\cal C}_2\Big)$.
To obtain the unitary representations of $\mathfrak{sl}(2,\mathbb C)$, we use the relation
\eqref{eq:lor}  combined with the substitution $z^1=z_+, z^2=z_-$.
Using this  parameterisation, the functions $\psi_{\ell_0,\ell_1}^{s,m}$  take the form
{\small
\beqa
\label{eq:harmsl2c}
\psi_{\ell_0,\ell_1}^{s,m}&=& A_{s}^{\ell_0,\ell_1} \sqrt{(2s+1)(s+m)!(s-m)!(s+\ell_0)!(s-\ell_0)!} \ \
(\cos \Theta\cos \bar \Theta e^{2 \p_{+1} }+\sin \Theta\sin \bar \Theta e^{2 \p_{-1}}  )^{\ell_1-s-1}
\times \nn\\
&&\sum \limits_k e^{i\big((m+\ell_0+k) \Phi_+ - k \bar \Phi_+
+(s-m-k) \Phi_- - (s-\ell_0 -k) \bar \Phi_-\big)}
\frac{\cos^{m+\ell_0+k} \Theta (-\cos \bar \Theta)^k \sin^{s-m-k} \Theta \sin^{s -\ell_0 -k} \bar \Theta}
      {(m+\ell_0+k)! k ! (s-m-k)! (s-\ell_0 -k)!}  \nn \ .
\eeqa
}

 We now define  a scalar product on 
${\mathbb S}^3_{\mathbb C} \setminus\Big({\cal C}_2 \times {\cal C}_2\Big)$ according to the following prescription: 
\beqa
\label{eq:scalS3C}
(f,g)&=&\frac1{(2\pi)^2}\int_{{\cal S}^3_{\mathbb C}  \setminus\Big({\cal C}_2 \times {\cal C}_2\Big)}  \cos \Theta \sin \Theta   \cos \bar \Theta \sin \bar \Theta
\dd \vt_0 \dd \vt_1 \dd \p_{+0}\dd  \p_{+1}\dd  \p_{-0} \dd \p_{-1}
\\
&& \hskip 1.5truecm \times
\bar f(\vt_0,\vt_1,\p_{+0}, \p_{+1}, \p_{-0}, \p_{-1})
g(\vt_0,\vt_1,\p_{+0}, \p_{+1}, \p_{-0}, \p_{-1})
\ . \nn
\eeqa
 It is important to observe that this scalar product is not suitable to be adapted to
 the $\psi-$functions, since in this case the integrals
 will generally be divergent (see {\it e.g.} \cite{st} for a suitable scalar product for this case).  This will however not constitute a serious
constraint, as we will show that the corresponding scalar product adapted to the suitable real form of 
$\mathbb S^3_\mathbb C
 \setminus\Big({\cal C}_2 \times {\cal C}_2\Big)$ 
will circumvent this difficulty.

\subsection{Harmonics of  $\mathbb S_\mathbb C^3 \setminus\Big({\cal C}_2\times
{\cal C}_2\Big)$}
Recall that for a   differentiable manifold ${\cal M}$  endowed with 
a  non-singular metric tensor $g_{ij}$, the Laplacian is defined by
\beqa
\Delta = \frac1{\sqrt{g}} \partial_i\big(\sqrt{g} g^{ij} \partial_j \big) \ , \nn
\eeqa
where $g = \det(g_{ij})$ and $g^{ij}$ is the inverse of $g_{ij}$.
A harmonic is simply  an eigenfunction of $\Delta$. 
The fact that the representations of the three-dimensional (real and complex) Lie groups
we will considered in this paper 
are harmonic on appropriate manifolds  naturally means that the Laplacian is proportional to
the Casimir operator (the case of $E_2$ is somehow particular, as we will see).

The differential realisation of $\mathfrak{sl}(2,\mathbb C)$ is of interest for several reasons. Firstly,
the representations   are obtained by functions
living naturally on  
$\mathbb S^3_\mathbb C \setminus\Big({\cal C}_2 \times {\cal C}_2\Big)$.
 In addition, the functions \eqref{eq:harmsl2c}  have a further interesting property, namely that  they are harmonic.

\medskip
\noindent From the relation $z_+ \bar z'_+ + z_-\bar z'_- =1$ we get
\beqa
\dd^2 s &=&\text{d} z_+ \text{d}\bar z'_+ + \text{d}z_- \text{d}\bar z'_-=
\dd^2 \Theta + \cos^2 \Theta  \ \text{d}^2 \Phi_++ \sin^2 \Theta  \ \text{d}^2 \Phi_- \ ,\nn\\
\text{d}^2 \bar s&=&\text{d}^2 \bar \Theta + \cos^2 \bar \Theta \ \text{d}^2 \bar \Phi_++ \sin^2 \bar\Theta \
\text{d}^2 \bar \Phi_-\nn \ .
\eeqa
Introducing the  metric tensor  $g_{ij}$ together with its  inverse $g^{ij}$  and $g = \det(g_{ij})$, we can define the
Laplacian
\beqa
\Delta &=& \frac1{\sqrt{g}}\partial_i(\sqrt{g} g^{ij} \partial_j) +
\frac1{\sqrt{\bar g}}\bar \partial_i(\sqrt{\bar g} \bar g^{ij} \bar \partial_j)\nn\\
&=& \ \ \ \frac1{\cos \Theta \sin \Theta} \partial_\Theta(\cos \Theta \sin \Theta \partial_\Theta)
+\frac1{\cos^2\Theta} \partial_{\Phi_+}^2 + \frac1{\sin^2\Theta} \partial_{\Phi_-}^2\nn \\
&&+\ \frac1{\cos \bar \Theta \sin \bar \Theta} \partial_{\bar \Theta}(\cos \bar \Theta \sin \bar \Theta \partial_{\bar \Theta})
+\frac1{\cos^2 \bar\Theta} \partial_{\bar\Phi_+}^2 + \frac1{\sin^2\bar \Theta} \partial_{\bar \Phi_-}^2 \ .\nn
\eeqa
A direct computation shows that the Laplacian reduces to  a Casimir operator, {\it i.e.}
\beqa
-\frac12 \Delta&=& L_0^2 + \frac12(L_+ L_- + L_- L_+) + \bar L_0^2 + \frac12(\bar L_+ \bar L_- + \bar L_- \bar L_+) \nn\\
      &=& \frac12\big[J_0^2 +\frac12(J_+ J_- +J_- J_+)\big] - \frac12\big[K_0^2 +\frac12(K_+ K_- +K_- K_+)\big] \ . \nn
\eeqa
 Therefore, for  a given monomial we have
\beqa
\Delta( z_+^a z_-^b\bar z_-^c \bar z_+^d) +
\Big((a+b) (a+b +2) +  (c+d)((c+d +2)\Big)  z_+^a z_-^b\bar z_-^c \bar z_+^d  =0 \ .\nn
\eeqa
 This implies  that the functions $\psi^{s,m}_{\ell_0,\ell_1}$ are harmonic and satisfy the constraint
\beqa
\Delta \psi^{s,m}_{\ell_0,\ell_1}+
2(\ell_0^2+\ell_1^2-1)  \psi^{s,m}_{\ell_0,\ell_1}
=0 \ . \nn
\eeqa

\section{The $\mathfrak{su}(2)$ algebra}\label{sec:su2}
The $\mathfrak{su}(2)$ algebra is the so-called compact
 real form of the $\mathfrak{sl}(2,\mathbb C)$ algebra,  where the latter is considered now
as a complex three-dimensional algebra. The results of the previous section can be directly applied to obtain
a differential realisation of $\mathfrak{su}(2)$ on  the  Lie group $SU(2)$ itself, that is, the three-sphere ${\mathbb S}^3$,  corresponding to a
real form of the complex unit three-sphere $\mathbb S^3_\mathbb C$  (corresponding to a real form of the Lie
group $SL(2,\mathbb C)$ ).

\subsection{Unitary representations of $\mathfrak{su}(2)$ on 
$\mathbb S^3 \setminus \Big(\mathbb S^1\times \mathbb S^1\Big)$ }\label{sec:unisu2}
The $SU(2)-$Lie group is the set of two-by-to  unitary complex matrices of determinant one
\beqa
\label{eq:U-su(2)}
U =\bpm \alpha&\beta \\-\bar\beta&\bar\alpha \epm \ ,
\alpha, \beta \in \mathbb C\ , \big|\alpha\big|^2 + \big|\beta\big|^2 =1 \ . 
\eeqa
If we consider the real form of ${\mathbb S}^3_\mathbb C$ parametrised by
\beqa
\label{eq:s3}
\Theta = \vt_0 =\theta\ ,
\Phi_+ = \p_{+0} = \p_+ \ ,
\Phi_- = \p_{-0}=\p_- \ ,
\eeqa
in the formul\ae \ of Section \ref{sec:S3C} we obtain
\beqa
z_+, z'_+ \to w_+ = \cos \theta e^{i\p_+} \ ,\
z_-, z'_- \to  w_- = \sin \theta e^{i\p_-}\ , \ 0\le \theta \le \frac{\pi}2, 0 \le \p_\pm  <2 \pi \ ,
\nn
\eeqa
from which we derive the identity
\beqa
\big|w_+\big|^2 + \big|w_-\big|^2 =1 \ . \nn
\eeqa
Note that the angle $\theta$ above  must belong to the interval  $[0,\pi/2]$ if the moduli of $w_+$ and $w_-$ are positive.
Thus $w_+$ and $w_-$ naturally parameterise the three-sphere.
Now, removing the two circles from $\mathbb S^3$ defined by
\beqa
\theta&=&0\ , \ \ \p_+ \in [0,2\pi[ \ , \nn \\
\theta&=&\frac{\pi}2\ , \ \ \p_- \in [0,2\pi[ \ , \nn
\eeqa
we have a bijection between ${\cal I}_3= ]0,\pi/2[ \times [0,2\pi[ \times [0,2\pi[$ and 
$\mathbb S^3 \setminus \Big({\mathbb S^1}\times {\mathbb S^1}\Big)$ such that
the direct application is continuous  whereas the reciprocal application is not. Note also that
the manifold $\mathbb S^3 \setminus \Big({\mathbb  S^1}\times {\mathbb S^1}\Big)$  is a real form of the manifold
$\mathbb S^3_\mathbb C \setminus \Big({\cal C}_2\times {\cal C}_2\Big)$ in the sense given in the introduction.

 With this parameterisation
the matrix elements of $U$ in \eqref{eq:U-su(2)} become 
  $\alpha=\cos \theta e^{i\p_+}$ and $\beta=\sin\theta e^{i\p_-}$. 
It is clear from this construction that our parameterisation of  $SU(2)$  differs from the usual one in terms 
of the Euler angles \cite{bvd,LOU}. 

\medskip
\noindent 
In the same real form, the generators of the $\mathfrak{su}(2)-$algebra are simply obtained,  substituting \eqref{eq:s3} into
\eqref{eq:liesl2c} and \eqref{eq:liesl2c-2} ({\it i.e.} $L_\pm= \bar L_\pm \to R_\pm, L_0= \bar L_0 \to R_0$)
\beqa
\label{eq:R}
R_+ &=& \frac12 e^{i(\p_+ - \p_-)} \Big(-i \tan \theta \partial_{\p_+}  + \partial_\theta-i \cot\theta \partial_{\p_-}\Big)
\ \ , \nn\\
R_-&=& \frac12 e^{i(\p_- -\p_+)} \Big(-i \tan \theta \partial_{\p_+}  - \partial_\theta-i \cot\theta \partial_{\p_-}\Big)
 \ ,
 \\
R_0&=& -\frac{i}2 \Big(\partial_{\p_+} - \partial_{\p_-} \Big) \ , \nn
\eeqa
and satisfying 
\beqa
\label{eq:rot}
\big[R_0,R_\pm\big] = \pm R_\pm \ \ , \big[R_+,R_-\big] = 2 R_0 \ .
\eeqa
The spinor representation is given by
\beqa
{\cal D}_\frac12 = \Big\{ w_+, w_-\Big\} \cong\Big\{ \bar w_-,\bar w_+\Big\}   \ , \nn
\eeqa
 whereas the finite dimensional representations are determined by
\beqa
\label{eq:DD}
{\cal D}_{\frac{\ell}2} = \Bigg\{\Phi_{\ell,m} &=& \sqrt{\frac{(2 \ell+1)!}{(\ell+m)!(\ell-m)!}} w_+^{\ell +m} w_-^{\ell-m}
\nn\\
&=&
\sqrt{\frac{(2 \ell+1)!}{(\ell+m)!(\ell-m)!}}
e^{i(\ell+m)\p_+ + i (\ell -m) \p_-} \cos^{\ell +m} \theta \sin^{\ell -m} \theta, -\ell \le m \le \ell \Bigg\} \ , 
\eeqa
with $\ell \in  \mathbb N$.
We have  the action
\beqa
R_+ \Phi_{\ell,m} &=& \sqrt{(\ell -m)(\ell +m+1)} \Phi_{\ell, m+1} \ ,\nn \\
R_- \Phi_{\ell,m} &=& \sqrt{(\ell +m)(\ell -m+1)} \Phi_{\ell, m-1} \ , \nn \\
R_0 \Phi_{\ell,m} &=& m \Phi_{\ell,m} \ . \nn
\eeqa
Introducing the $SU(2)-$invariant scalar product on the three-sphere  (or more precisely on $\mathbb S^3 
 \setminus\Big(\mathbb S^1\times \mathbb S^1\Big)$), 
which can be naturally obtained from the scalar product \eqref{eq:scalS3C}
using the corresponding real form \cite{WA}).
\beqa
\label{eq:scal-S3}
(f,g)= \frac1{2 \pi^2} \int\limits_0^{\frac{\pi}2} \text{d} \theta \int \limits_0^{2 \pi} \cos \theta \text{d} \p_+
 \int \limits_0^{2 \pi} \sin \theta \text{d} \p_- \bar g(\theta,\p_+,\p_-) f(\theta,\p_+,\p_-)  \ , 
\eeqa
 the orthogonality relation
\beqa
(\Phi_{\ell,m},\Phi_{\ell',m'}) = \delta_{\ell \ell'} \delta_{m m'} \  , \nn
\eeqa
 can be easily shown,  as well as the fact that the operators satisfy $R_\pm^\dag = R_\mp, R_0^\dag = R_0$ with respect to the scalar product.
 As a consequence,  the representation ${\cal D}_{\frac{\ell}2}$ is unitary.  We emphasise again 
that the Haar measure of $SU(2)$ considered here is different from the usual Haar measure, as follows from the choice 
of parameterisation taken above.
 Similar relations to these, using a slightly different formulation, are well known and were originally obtained in \cite{Beg}.
\subsection{Harmonics on $\mathbb S^3\setminus \Big(\mathbb S^1\times \mathbb S^1\Big)$}
For the considered real form, the  previous metric becomes
\beqa
\text{d}^2 s = \text{d}^2 \theta + \cos^2 \theta \ \text{d}^2 \p_+ + \sin^2 \theta \ \text{d}^2 \p_- \ ,\nn
\eeqa
while the  Laplacian on ${\mathbb S}^3 \setminus \Big(\mathbb S^1\times \mathbb S^1\Big)$ reduces to
\beqa
\Delta=\frac1{\cos \theta \sin \theta} \partial_\theta(\cos \theta \sin \theta \partial_\theta)
+\frac1{\cos^2\theta} \partial_{\p_+}^2 + \frac1{\sin^2\theta} \partial_{\p_-}^2\nn \ .
\eeqa
A  routine computation shows that the  Laplacian is related to the Casimir operator  by
\beqa
\Delta = - 4 \big(J_0^2 + \frac12(J_+ J_- + J_- J_+)\big)= -4 Q \ . \nn
\eeqa
 It follows at once  that the functions $\Phi_{\ell,m}$ are harmonic:
\beqa
\Delta \Phi_{\ell,m} + 4 \ell(\ell+1) \Phi_{\ell,m} = 0 . \nn
\eeqa

We now compare our approach with the standard approach.
It is well known that unitary representations can be obtained introducing two complex variables $z^1$ and $z^2$. Within this parameterisation the $\mathfrak{su}(2)$ generators
take the form \cite{wyb, ed}
\beqa
\label{eq:stant-gene}
R_+ &=& z^1 \partial_2 - \bar z_2 \bar \partial^1 \ , \nn\\
R_+ &=& z^2 \partial_1 - \bar z_1 \bar \partial^2 \ , \\
R_0&=& \frac12\Big(z^1 \partial_1 -z^2 \partial_2\Big)- \frac12\Big(\bar z_1  \bar \partial^1 -\bar z_2 \bar \partial^2\Big)\ , \nn
\eeqa
and unitary representations are given by
\beqa
\label{eq:stant-rep}
{\cal D_\ell} = \Bigg\{ \phi_{\ell,m}(z)=\frac1 {\sqrt{(\ell+m) ! (\ell -m) !}} (z^1)^{\ell +m} (z^2)^{\ell -m} \ , \ \ -\ell \le m \le \ell \Bigg\} \ , 
\eeqa
with $2 \ell \in \mathbb N$. 
The scalar product is given by
\beqa
\label{eq:stant-ps}
(f,g) = -\frac 1{4\pi^2} \int_{\mathbb C^2} \text{d}^2 z^1 \text{d}^2 z^2  \bar f(\bar z) g(z) e^{-|z^1|^2 -|z^2|^2} \ ,  
\eeqa
with d${}^2z =$d$z$d$\bar z= 2ir$d$r$d$\theta$ in polar coordinates,
and we have
\beqa
(\phi_{\ell,m}, \phi_{\ell',m'}) = \delta_{\ell \ell'} \delta_{m m'} \ . \nn
\eeqa
If we now substitute
\beqa
z^1 &=&\cos \theta_1 e^{i \varphi_1}\ , \nn\\
z^2 &=& \sin \theta_2 e^{i \varphi_2} \ , \nn
\eeqa
in \eqref{eq:stant-gene},
we obtain
\beqa
\label{eq:gene2}
R_+ &=& e^{i\big(\varphi_2 -\varphi_1\big)} \Bigg( \frac{\sin \theta_2}{\sin \theta_1} \partial_{\theta_1} + \frac{\cos \theta_1}{\cos \theta_2} \partial_{\theta_2} -
i \frac{\sin \theta_2}{\cos \theta_1} \partial_{\varphi_1}  - i \frac{\cos \theta_1}{\sin \theta_2} \partial_{\varphi_2} \Bigg) \ , \nn \\
R_- &=& e^{i\big(\varphi_1 -\varphi_2\big)} \Bigg( -\frac{\sin \theta_2}{\sin \theta_1} \partial_{\theta_1} - \frac{\cos \theta_1}{\cos \theta_2} \partial_{\theta_2} -
i \frac{\sin \theta_2}{\cos \theta_1} \partial_{\varphi_1}  - i \frac{\cos \theta_1}{\sin \theta_2} \partial_{\varphi_2} \Bigg) \ ,  \\
R_0&=&-i\Big(\partial_{\varphi_1} - \partial_{\varphi_2}\Big) \nn \ . 
\eeqa
To take the limit   $\theta_1 = \theta_2 = \theta$ some care is needed. Indeed, we cannot simply substitute $\theta_1=\theta$ and  $\theta_2=\theta$ in the formul\ae \ above, in
particular, we have to treat correctly the  derivatives $\partial_{\theta_1}$ and  $\partial_{\theta_2}$  whenever encountered. For instance,  the following facts must be observed:
\beqa
\label{eq:limm}
&&\left.
\begin{array}{l}
\displaystyle \lim_{\begin{array}{l}\theta_1 \to \theta\\ \theta_2 \to \theta\end{array}}  \partial_{\theta_1}  \ , \\ \\
\displaystyle \lim_{\begin{array}{l}\theta_1 \to \theta\\ \theta_2 \to \theta\end{array}}  \partial_{\theta_2} \ , \\
\end{array}\right\} \ \ \mbox{do not exist} \\ \nn
&&\hskip .25truecm \displaystyle \lim_{\begin{array}{l}\theta_1 \to \theta\\ \theta_2 \to \theta\end{array}} \Big( \partial_{\theta_1}+\partial_{\theta_2}\Big) = \partial_\theta \ .
\eeqa
This can be shown easily.  Using \eqref{eq:stant-rep}, it follows that all functions $f(\theta_1,\theta_2)$ are written as a sum of  products of the type $f(\theta_1,\theta_2)= g_1(\theta_1) g_2(\theta_2)$.
Thus
\beqa
\displaystyle \Bigg[\lim_{\begin{array}{l}\theta_1 \to \theta\\ \theta_2 \to \theta\end{array}}
\Big( \partial_{\theta_1}+\partial_{\theta_2}\Big)\Big( g_1(\theta_1) g_2(\theta_2)\Big)\Bigg]=
\partial_\theta g_1(\theta) g_2(\theta) + g_1(\theta) \partial_\theta g_2(\theta) = \partial_\theta\Big(g_1(\theta) g_2(\theta)\Big)\ , \nn
\eeqa
legitimating the limit of $\partial_{\theta_1} + \partial_{\theta_2}$ in \eqref{eq:limm}.  On the contrary
\beqa
\displaystyle \Bigg[\lim_{\begin{array}{l}\theta_1 \to \theta\\ \theta_2 \to \theta\end{array}}  \partial_{\theta_1} g_1(\theta_1) g_2(\theta_2)\Bigg]=
\Big(\partial_\theta g_1(\theta)\Big) g_2(\theta) \ ,
\eeqa
which cannot be obtained as a limit, {\it i.e.}, by a definite operator acting on $g_1(\theta) g_2(\theta)$.

Thus taking the limit $\theta_1=\theta_2 = \theta$, using \eqref{eq:limm} in \eqref{eq:gene2} reproduces  \eqref{eq:R}. Moreover
\eqref{eq:stant-rep}  reduces to  \eqref{eq:DD} and \eqref{eq:stant-ps} restricted to the sphere $\mathbb S^3$  reproduces the scalar product \eqref{eq:scal-S3}. The overall factor 
in \eqref{eq:DD}, in the corresponding limit, must be added in order to have normalised functions.  This means that
the real form considered above can be obtained from the standard approach in the limit \eqref{eq:limm}. 
Of course, any other parameterisation of $z^1, z^2$ would give rise to a different but equivalent explicit realisation of unitary representation
of $\mathfrak{su}(2)$.

If we  complexify the  relations above  for $\mathfrak{su}(2)$, we get the
$\mathfrak{sl}(2,\mathbb C)$ generators given in the previous section.
In particular, if  we make the following
substitution
\beqa
\theta &\longrightarrow &\Theta = \vartheta_0 + i \vartheta_1 \ , \nn \\
\varphi_1  &\longrightarrow &\Phi_1 = \varphi_{1,0} + i \varphi_{1,1} \ , \nn \\
\varphi_2  &\longrightarrow &\Phi_2 = \varphi_{2,0} + i \varphi_{2,1} \ , \nn \\
\eeqa
\eqref{eq:R}  reduces to \eqref{eq:liesl2c} and the complex conjugated substitution (with $\theta\to\bar \Theta,
\varphi_1\to\bar \Phi_1$ and
$\varphi_2\to\bar \Phi_2$) leads to \eqref{eq:liesl2c-2}. This is natural since the complexification of $\mathfrak{su}(2)$
gives rise to $\mathfrak{sl}(2,\mathbb C)$ and this procedure is the  ``opposite'' procedure (real form) which
gives the $\mathfrak{su}(2)$ generators from the  $\mathfrak{sl}(2,\mathbb C)$ generators.

\section{The $\mathfrak{sl}(2,\mathbb R)$ algebra}
The group $SL(2,\mathbb R)$ is the group of two-by-two real unimodular  matrices.  By means of the unitary matrix
\beqa
A=\frac{1}{\sqrt{2}}\left(
\begin{array}{cr}
1 & -i\\
1 & i
\end{array}
\right) \ ,
\eeqa
it can be easily shown that this group is isomorphic
to the group $SU(1,1)$ defined by
\beqa
U=\bpm \alpha&\beta \\ \bar\beta&\bar\alpha\epm\ , \ \ \alpha,\beta\in \mathbb C\ , \ \
\big|\alpha\big|^2-\big|\beta\big|^2=1 \ . \nn
\eeqa
The isomorphism is given by the relation (see {\it e.g.} \cite{Ino})
\beqa
SU(1,1)=A\; SL(2,\mathbb{R})\; A^{\dagger}\ .
\eeqa
It turns out that the Lie group $SU(1,1)$ can be identified as topological space with the hyperboloid $\mathbb H_{2,2}$ of signature $(2,2)$.

\medskip
\noindent We denote by  $J_\pm, J_0$
the $\mathfrak{sl}(2,\mathbb R) \cong \mathfrak{su}(1,1)$-generators   that  satisfy  the commutators
\beqa
[J_0, J_\pm] = \pm J_\pm\ , \ \ [J_+, J_-] = -2 J_0 \ . \nn
\eeqa
 Over this basis,  the Casimir operator is given by
\beqa
Q= J_0^2 -\frac12(J_+ J_- + J_- J_+) \ . \nn
\eeqa

\subsection{Unitary representations}\label{sec:sl2Rrep}
Since $SL(2,\mathbb R)$ is a non-compact group, its unitary representations are infinite dimensional.
They have largely been studied by different authors, and classifications can be found in \cite{bar,ggv,lang,wyb,UI}.
 Representations of the first type are either bounded from  below or  bounded from above,
corresponding  to the so-called discrete Bargmann series \cite{bar}:

\beqa
\label{eq:disc} {\cal D}_s^+&&\left\{
\begin{array}{lll}
J_+ \big|s_+,n\big>& =&\sqrt{(n+1)(n+2 s)} \big|s_+,n+1\big>  \ ,\\
J_- \big|s_+,n\big> &=&\sqrt{n(n+2 s-1)} \big|s_+,n-1\big>  \ , \\
J_0 \big|s_+,n\big> &=& (n+s)  \big|s_+,n\big>  \ , \\
Q \big|s_+,n\big> &=& s(s-1) \big|s_+,n\big>  \ ,
\end{array}\right.\nn \\\\
{\cal D}_s^-&&\left\{
\begin{array}{lll}
J_+ \big|s_-,n\big> &=&-\sqrt{n(n+2 s-1)} \big|s_-,n-1\big>  \ , \\
J_- \big|s_-,n\big>& =&-\sqrt{(n+1)(n+2 s)} \big|s_-,n+1\big>  \ , \\
J_0 \big|s_-,n\big> &=& -(n+s)  \big|s_-,n\big>  \ , \\
Q \big|s_-,n\big> &=& s(s-1) \big|s_-,n\big>  \ ,
\end{array} \right. \nn
\eeqa
where $s \in \mathbb R$. It can be shown that these representations are finite-dimensional whenever $s$
is a negative half-integer number. Since $n \in \mathbb N$, the representation ${\cal D}_s^+$ is
bounded from below and the representation ${\cal D}_s^-$ bounded
from above.  If we perform the substitution
$J_0 \to -J_0, J_\pm \to -J_\mp$, it can be shown that these two representations are isomorphic.
In addition, they are unitary and can be  exponentiated  when $s>0$.

\noindent  Representations of the second type are neither bounded from below nor above but unbounded.
They correspond to the continuous series
\beqa
\label{eq:cont}
{\cal D}_{\lambda,\mu} \ \ \left\{
\begin{array}{lll}
J_+ \big| \lambda,\mu,n\big> &=& \sqrt{(2\mu +n+1)(n-2\lambda)}  \big| \lambda,\mu,n+1\big> \ ,  \\
J_- \big| \lambda,\mu,n\big> &=& \sqrt{(2\mu +n)(n-1-2\lambda)}  \big| \lambda,\mu,n-1\big> \ ,  \\
J_0 \big| \lambda,\mu,n\big> &=& (n-\lambda+\mu) \big| \lambda,\mu,n-1\big> \ ,  \\
Q  \big| \lambda,\mu,n\big> &=& (\lambda+\mu) (\lambda+\mu+1)\big| \lambda,\mu,n-1\big> \ ,
\end{array}
\right.
\eeqa
with $\mu, \nu \in \mathbb C$ and $n \in \mathbb Z$. The representations are unitary whenever  all operators are Hermitean.
 As a consequence,
the eigenvalues of $J_0$ are real, and  $\mu -\lambda \in \mathbb R$.
If we set $\lambda + \mu = \Phi_0 + i \Phi_1$,  the Casimir operator reduces to
\beqa
Q= \Phi_1^2 + \Phi_1 -\Phi_2^2 + i\Phi_2(2 \Phi_1 +1) \ . \nn
\eeqa
 We observe that the representations ${\cal D}_{\lambda,\mu}$ and
${\cal D}_{(\lambda+\frac12), (\mu-\frac12)}$ are isomorphic, thus we can restrict  ourselves to the case $ -1/2< \mu -\lambda\le 1/2$.
Two continuous representations must be distinguished:
 \begin{itemize}
\item  the continuous principal series
\beqa
\label{eq:cont1}
 -1/2< \mu -\lambda\le 1/2 \ , \ \ \Phi_1 = -\frac12 \ , \ \ \text{and} \  \ \Phi_2 = \sigma >0 \ ,
\eeqa
with
\beqa
Q= -\frac14 -\sigma^2 < -\frac14 \ . \nn
\eeqa
\item the continuous supplementary series
\beqa
\label{eq:cont2}
-1/2< \mu -\lambda\le 1/2\ ,\ \  \Phi_2=0 \ , \ \  \Phi_1=\lambda + \mu \in \mathbb R \ , \ \ \text{with} \ \
|\lambda+\mu +\frac12|<\frac12 -| \lambda-\mu| \ .
\eeqa

 \end {itemize}
 Observe that, in contrast to the $\mathfrak{sl}(2,\mathbb C)$ or $\mathfrak{su}(2)$ Lie algebras,
the spin can adopt an arbitrary real value. This is due to the fact that
the first homotopy group of $SU(1,1)$ is  isomorphic to  $\mathbb Z$,  as well as that the  representations
considered are defined on some appropriate covering group of $SU(1,1)$ (see  {\it e.g.} \cite{NA} for details).

\subsection{Realisation of Unitary  representations of $\mathfrak{sl}(2,\mathbb R)$  on
$\mathbb H_{2,2}\setminus \mathbb S^1$ }\label{sec:repsl2}

\noindent The Lie algebra $\mathfrak{su}(1,1)$ is a real form of
the three-dimensional complex Lie algebra $\mathfrak{sl}(2,\mathbb
C)$. Correspondingly, we now show that all its unitary
representations can be obtained on a appropriate covering of a
real form of $\mathbb{S}^3_\mathbb C \setminus\Big({\cal C}_2 \times {\cal C}_2 \Big)$.
 As observed before,  we can identify the Lie group $SU(1,1)$ with the hyperboloid
$\mathbb H_{2,2}$ of signature $(2,2)$, which is
a real form of the Lie group $SL(2,\mathbb C)$, {\it i.e.}, the unit complex three-sphere ${\cal S}^3_\mathbb C$ that we parameterise using
\beqa
\label{eq:su1,1}
\Theta = i\vt_1 = i\rho\ , \ \Phi_+=  \p_{+0} = \p_+ \ , \Phi_-= \phi_{-0} = \p_- \ ,\nn
\eeqa
in the formul\ae \  of Section \ref{sec:S3C}, the complex numbers $z_\pm$ and $z'_\pm$  reducing to
(we have multiplied $\zeta_-$ by $-i$ for convenience)
\beqa
\label{eq:H22}
 \zeta_+= \cosh\rho e^{i\p_+} \ , \ \zeta_-= \sinh \rho e^{i\p_-} \ , \ \rho \in \mathbb R_+\ ,
0\le \p_\pm < 2 \pi \ ,
\eeqa
which obviously satisfy
\beqa
\big| \zeta_+\big|^2-\big| \zeta_-\big|^2 =1 \ . \nn
\eeqa
The variables $\zeta_\pm$ clearly parameterise the hyperboloid $\mathbb H_{2,2}$.
If we now remove one circle from $\mathbb H_{2,2}$
\beqa
\rho=0 \ , \ \ \p_+ \in [0,2\pi[ \ , \nn
\eeqa
 we obtain a bijective map between ${\cal I}_{2,2}= [0,2\pi[\times [0,2\pi[ \times \mathbb R_+^*$ and $\mathbb H_{2,2}\setminus \mathbb S^1$ such that
the direct application is  continuous,  but the reciprocal application being not continuous.
Note also that the manifold $\mathbb H_{2,2}\setminus \mathbb S^1$ is a real form of the manifold
$\mathbb S^3_\mathbb C \setminus\Big({\cal C}_2 \times {\cal C}_2\Big)$ in the sense given of the introduction.
Now it can be shown that the manifold $]1,+\infty[ \times \mathbb S^1 \times \mathbb S^1$ is homeomorphic   to 
$\mathbb H_{2,2}\setminus \mathbb S^1$. Indeed, 
\beqa
\begin{array}{llll}
f:&]1,+\infty[ \times \mathbb S^1 \times \mathbb S^1&\to&\mathbb H_{2,2}\setminus \mathbb S^1\\
&(r,u_+,u_-)&\mapsto&\left\{\begin{array}{l}
                           \zeta_+ = r u_+ \\
                           \zeta_- = \sqrt{r^2-1}u_-
                           \end{array}\right.
\end{array}\nn
\eeqa
is clearly bijective. Recalling that $\mathbb H_{2,2}\setminus \mathbb S^1$
is given by $|\zeta_+|^2-|\zeta_-|^2 =1$ with $|\zeta_+|>1$, the reciprocal application is defined by
\beqa
\begin{array}{llll}
f^{-1}:&\mathbb H_{2,2}\setminus \mathbb S^1&\to&]1,+\infty[ \times \mathbb S^1 \times \mathbb S^1\\
&(\zeta_+, \zeta_-)&\mapsto&\left\{\begin{array}{l}
                           r = | \zeta_+| \\
                           u_+ = \frac{\zeta_+}{|\zeta_+|}\\
                           u_-=\frac{\zeta_-}{\sqrt{|\zeta_+|^2-1}}
                           \end{array}\right.
\end{array}\nn
\eeqa
It is routine to show that $f$ and $f^{-1}$ are continuous. 
Since $]1,+\infty[$ is contractible,  $\pi_1\Big(]1,+\infty[\times \mathbb S^1 \times \mathbb S^1\Big)
=\pi_1\Big( \mathbb S^1 \times \mathbb S^1\Big)=  \mathbb Z \times \mathbb Z$ and we have
$\pi_1(\mathbb H_{2,2}\setminus \mathbb S^1) = \mathbb Z \times \mathbb Z$.
 This enables us to determine appropriate coverings of $\mathbb H_{2,2}\setminus \mathbb S^1$
which are defined  in dependence of the parameterisation:
\begin{enumerate}
\item[-] the $(p_+,p_-)-$sheeted covering 
$\widetilde{\mathbb H_{2,2}\setminus \mathbb S^1}^{(p_+,p_-)}$  parameterised by 
$\rho \in \mathbb R_+^*, \ 0\le \p_\pm < 2 p_\pm \pi$
\item[-] the $(\infty,p_-)-$sheeted  covering  
$\widetilde{\mathbb H_{2,2}\setminus \mathbb S^1}^{(\infty,p_-)}$  parameterised by 
  $\rho \in \mathbb R_+^*, \p_+ \in \mathbb R, \  0\le \p_- < 2 p_- \pi$
\item[-] the $(p_+,\infty)-$sheeted  covering
$\widetilde{\mathbb H_{2,2}\setminus \mathbb S^1}^{(p_+,\infty)}$  parameterised by 
 $\rho \in \mathbb R_+^*, \ \p_- \in \mathbb R, \  0\le \p_+ < 2 p_- \pi$
\item[-] the $(\infty,\infty)-$sheeted  covering
$\widetilde{\mathbb H_{2,2}\setminus \mathbb S^1}^{(\infty,\infty)}$  parameterised by 
 $\rho \in \mathbb R_+^*, \ \p_+,\p_- \in \mathbb R.$
\end{enumerate}

\noindent Next, in the same manner as for $\mathfrak{su}(2)$, the $\mathfrak{sl}(2,\mathbb C)$-generators reduce to
\beqa
\label{eq:sl2H22}
J_+&=&\frac12 e^{i(\p_+ - \p_-)}\Big(-i \tanh(\rho) \partial_{\p_+} - \partial_\rho +i\coth \rho \partial_{\p_-}\Big)\ , \nn \\
J_-&=&\frac12 e^{i(\p_- - \p_+)}\Big(-i \tanh(\rho) \partial_{\p_+} + \partial_\rho +i\coth \rho \partial_{\p_-}\Big)  \ ,\\
J_0&=&-\frac{i}2 \Big(\partial_{\p_+} -\partial_{\p_-} \Big) \ , \nn
\eeqa
leading to
\beqa
\Big[J_0,J_\pm\Big] = \pm J_\pm \ , \ \ \Big[J_+,J_-\Big] = -2 J_0 \ . \nn
\eeqa
Note that in order to reproduce the usual commutation relation, such that for unitary representations we have
$J_\pm^\dag = J_\mp$, in the substitution above  we have multiplied $J_\pm$ by a factor $-i$ (see below).

The Casimir operator is given by
\beqa
Q=J_0^2-\frac12\big(J_+J_- +J_-J_+\big) \ . \nn
\eeqa

\noindent Interestingly, the differential realisation \eqref{eq:sl2H22} which defines 
 a left action of $\mathfrak{sl}(2,\mathbb R)$  on  $\mathbb H_{2,2} \setminus \mathbb S^1 \subset
SL(2,\mathbb R)$ extends on an appropriate covering   of
$\mathbb H_{2,2} \setminus \mathbb S^1$.
Furthermore, these definitions enable us to obtain   explicit  realisations of
 unitary representations of either the $p-$sheeted covering
or the universal covering of $SL(2,\mathbb R)$.

\medskip
\noindent 
As the spinor representation given by
\beqa
{\cal D}_\frac12 = \Big\{\zeta_+, \zeta_-\Big\} \cong\Big\{\bar \zeta_+, \bar \zeta_-\Big\}  \ , \nn
\eeqa
lives on $\mathbb {H}_{2,2} \setminus \mathbb S^1$, all unitary representations can be defined on 
$\mathbb H_{2,2}\setminus \mathbb S^1$ (or
one of its coverings).
We  can in principle obtain all representations of $\mathfrak{su}(1,1)$ with either $(\zeta_+, \zeta_-)$
or $(\bar \zeta_-, \bar \zeta_+)$ but, depending  on the representation considered, only one choice
would be consistent
with the scalar product on  $\mathbb H_{2,2}\setminus \mathbb S^1$  defined  below.
The unitary representation of  Section \ref{sec:sl2Rrep} can be naturally defined on $\mathbb H_{2,2}
\setminus \mathbb S^1$
(eventually on some $p$-sheeted covering  or even on its universal  covering).

\medskip
\noindent
For the discrete series bounded from below  we define
\beqa
{\cal D}_s^+ = \Bigg\{ \Psi_{s,n}^+ &=& \sqrt{\frac{2 \Gamma(n+2s)}{\Gamma(-1+2s)\Gamma(n+1)}}\bar \zeta_+{}^{-2s-n}
\bar\zeta_-{}^n
\nn \\
 &=& \sqrt{\frac{2 \Gamma(n+2s)}{\Gamma(-1+2s)\Gamma(n+1)}}
e^{-in\p_-+i(2s+n) \p_+ } \cosh^{-2s-n} \rho \sinh^{n}\rho \ , \ \ n \in \mathbb N \Bigg\} \ , \nn
\eeqa
while for the  discrete series bounded from above
\beqa
{\cal D}_s^- = \Bigg\{ \Psi_{s,n}^- &=& \sqrt{\frac{2 \Gamma(n+2s)}{\Gamma(-1+2s)\Gamma(n+1)}} \zeta_+^{-2s-n} \zeta_-^n
\nn \\
 &=& \sqrt{\frac{2 \Gamma(n+2s)}{\Gamma(-1+2s)\Gamma(n+1)}}
e^{- i(2s+n) \p_+ +in\p_-} \cosh^{-2s-n} \rho \sinh^{n}\rho \ ,\ \  n \in \mathbb N \Bigg\} \ , \nn
\eeqa
with $s>0$.
Unbounded representations and the continuous series are defined by
\beqa
{\cal D}_{\lambda,\mu}= \Bigg\{\Psi_{\lambda,\mu,n} &=& \sqrt{\frac{2 \Gamma(-2\lambda+n)}{\Gamma(2 \mu +n+1)(
\Gamma(-2\mu -2 \lambda-1)}} \bar\zeta_+{}^{2 \lambda -n} \bar\zeta_-{}^{2\mu +n} \nn \\
 &=& \sqrt{ \frac{2\Gamma(-2\lambda+n)}{\Gamma(2 \mu +n+1)(
\Gamma(-2\mu -2 \lambda-1)}}  \nn \\
&& \hskip 3.truecm \times \ e^{-i(2\lambda-n)\p_+ -i(2\mu +n) \p_-}  \cosh^{2 \lambda -n}\rho \sinh^{2 \mu +n} \rho \ , \ \ n\in \mathbb Z
\Bigg\} \ .\nn
\eeqa
Note that we could have defined the continuous series with $\zeta_\pm$ instead of $\bar \zeta_\pm$.  Both possibilities
however lead to identical conclusions. Unitarity is ensured if the parameters  
$\lambda,\mu$ are given by \eqref{eq:cont1} or \eqref{eq:cont2}.
If $s$, or $\lambda,\mu$ are not integers, we have to define the  $\Psi-$functions on a  suitable covering
space  of $\mathbb H_{2,2}\setminus \mathbb S^1$. 
In particular, in order that the formul\ae \  above make sense, the representations must be defined on some covering  of
$\mathbb H_{2,2} \setminus \mathbb S^1$.
For the discrete series,  for $ 2s=p/q \in \mathbb Q$, the representations are defined on the $(q,1)-$sheeted covering
of $\mathbb H_{2,2} \setminus \mathbb S^1$, whereas for   
 $s \in \mathbb R \setminus \mathbb Q$, the representations are defined on the $(\infty,1)-$sheeted covering
of $\mathbb H_{2,2} \setminus \mathbb S^1$.
For the continuous series, the representations are defined respectively on the $(p,p'),(\infty,p'),(p,\infty), (\infty,\infty)-$sheeted 
covering of $\mathbb H_{2,2} \setminus \mathbb S^1$,
where   $(\text{Re}(\mu), \text{Re}(\lambda))$  are  respectively (rational, rational), (irrational, rational), (irrational, rational),
(irrational, irrational)  with $\text{Re}(z)$ the real part of $z$.

\medskip
\noindent
We can now introduce an $SU(1,1)-$invariant scalar product appropriately adapted to the  discrete series
on the $(q, 1)-$covering  or the $(\infty, 1)-$covering   of $\mathbb H_{2,2} \setminus \mathbb S^1$
(which can be naturally obtained from the scalar product \eqref{eq:scalS3C}
using the corresponding real form),  respectively:
\beqa
\label{eq:scalhyper}
(f,g)_{(q,1)} &=& \frac1{q}\frac1{(2 \pi)^2}
\int \limits_0^{+\infty} \cosh \rho \sinh\rho \ \text{d} \rho \int \limits_0^{2 q \pi}
\text{d} \p_+ \int \limits_0^{2  \pi} \text{d} \p_-  \bar f(\rho,\p_+,\p_-)  g(\rho,\p_+,\p_-)  \ ,\nn\\
(f,g)_{(\infty,1)} &=& \frac 2{(2 \pi)^2}
\int \limits_0^{+\infty} \cosh \rho \sinh\rho \text{d} \rho \int \limits_{-\infty}^{+ \infty}
\text{d} \p_+ \int \limits_0^{2  \pi} \text{d} \p_-  \bar f(\rho,\p_+,\p_-)  g(\rho,\p_+,\p_-)  \ .\nn
\eeqa
For the discrete series,  
the integral involving the $\Psi-$functions can be computed easily when $s$ is a rational number, first performing the change of variables
$\cosh \rho =r$  leading to integrals of the form
\beqa
I_{a,b} =  \int \limits_1^{+\infty} r^{2a+1}(r^2-1)^b \text{d} r \ . \nn
\eeqa
The latter integrals are related to   hypergeometric functions \cite{SZE},  and one can show
that $I_{a,b}$  is convergent
for $a,b \in \mathbb R $  satisfying the inequalities $(a+b)<-1, b > -1$ (which automatically implies that $-a>0$):
\beqa
I_{a,b} =  \frac12\frac{\Gamma(1+b) \Gamma(-a-b-1)}{\Gamma(-a)} \ .\nn
\eeqa
Applying these results to the unitary representation of $\mathfrak{sl}(2,\mathbb R)$ constructed
on $\mathbb H_{2,2}\setminus \mathbb S^1$,  we have the following results.
For the discrete series bounded from above and below, the integrals converges if $s>1/2$, leading to
(for $s,s'>1/2$ and $s  = p/(2q) ,s' = p'/(2q')\in \mathbb Q$)
\beqa
(\Psi^\epsilon_{s,m}, \Psi^{\epsilon'}_{s',m'})_{( q'' ,1)} = \delta_{\epsilon \epsilon'} \delta_{ss'}\delta_{mm'} \ , \nn
\eeqa
with $q''$ the  least common multiple of $q,q'$.
Since $J_\pm^\dag = J_\mp, J_0^\dag = J_0$ for the scalar product,
 the representations is unitary if $s>1/2$. 

\medskip
\noindent 
Whenever $s$ or $s'$ is a irrational number, still in the case of the discrete series, 
using the integral representation of the Dirac $\delta-$distribution implies the 
identity

\beqa
(\Psi^\epsilon_{s,m}, \Psi^{\epsilon'}_{s',m'})_{(\infty,1)} = \delta_{\epsilon \epsilon'} \delta(s-s')\delta_{mm'} \ . \nn
\eeqa

If we proceed along the same lines as for the unbounded representation, {\it i.e.}, defining 
 a scalar product on an appropriate covering of $\mathbb H_{2,2}\setminus \mathbb S^1$,
 it turns out that the corresponding integrals diverge. This obstruction can be surmounted by considering an adapted scalar product, like that defined in \cite{ggv,nai}.

\subsection{ Harmonics  of $\mathbb H_{2,2}\setminus \mathbb S^1$}
For the real form $\mathbb H_{2,2} \setminus \mathbb S^1$ of 
${\cal S}^3_\mathbb C \setminus\Big({\cal C}_2 \times {\cal C}_2\Big)$ the metric becomes
\beqa
\text{d}^2 s = -\text{d}^2 \rho + \cosh^2 \rho \ \text{d}^2 \p_+ -\sinh^2\rho \ \text{d}^2 \p_- \ . \nn
\eeqa
A simple computation shows that the Laplacian is related to the Casimir operator
\beqa
\label{eq:lapsl2}
\Delta = -\frac1{\cosh \rho \sinh \rho} \partial_\rho (\cosh \rho \sinh \rho \partial_\rho) +
\frac1{\cosh^2 \rho} \partial^2_{\p_+} - \frac1{\sinh^2 \rho} \partial^2_{\p_-}  =  -4 Q\ .
\eeqa
In particular we have
\beqa
\Delta \Psi^\pm_{s,m} +4s (s-1)   \Psi^\pm_{s,m}=0 \ , \ \
\Delta \Psi_{\lambda,\mu,m}+4(\lambda+\mu)(\lambda+\mu+1) \Psi_{\lambda,\mu,m}=0 \ , \nn
\eeqa
 hence the $\Psi-$functions are harmonic.

 In order to recover the standard approach, we proceed as in Section \ref{sec:su2} and realise the $\mathfrak{sl}(2,\mathbb R)$ algebra in terms of two complex variables.
Following \cite{wyb}, as done for the $\mathfrak{su}(2)$ algebra, a differential realisation of $\mathfrak{sl}(2,\mathbb R)$ can be
 obtained in terms of two complex variables
$w^1,w^2$:
\beqa
\label{eq:stand-sl2}
J_+&=& -w^1 \partial_2 - \bar w_1 \bar \partial^2 \ , \nn\\
J_-&=& w^2 \partial_1 +\bar w_2 \bar \partial^1 \ , \\
J_0 &=&\frac12\Big(w^1 \partial_1 - w^2 \partial_2\Big) - \frac12\Big(\bar w_1\bar \partial^1 -\bar w_2 \bar\partial^2\Big) \ . \nn 
\eeqa
With this realisation one can explicitly construct all unitary representations of $\mathfrak{sl}(2,\mathbb R)$ \cite{wyb}.
For instance the representations bounded from below are given by
\beqa
{\cal D}^+_s = \Bigg\{ \psi_{s,m}^+ (w) = \sqrt{\frac{\Gamma(2s+m)}{\Gamma(m+1)}} \bar w_1^{-2s -m} \bar w_2^n\ , \ \ m \in \mathbb N \Bigg\} \ , \nn
\eeqa
with $s>0$.
Similar expressions for the other unitary representations hold \cite{wyb}.
To obtain a scalar product in this case is more delicate for at least two reasons: (i) the integral will in general not converge, (ii)  due to the simple connectedness of $\mathbb C^2$, representations
corresponding to representations in some covering space of $SL(2,\mathbb R)$ are generally not defined. There are several ways to circumvent this problem, for instance defining representations
on the half-plane, or on the unit disc.  This formulation has been used for example in \cite{ggv,lang}.  There exists, however, an alternative  way to proceed, and directly deducible from the previous analysis.
Indeed, if we now set
\beqa
\label{eq:sl-w}
w^1 &=& \cosh \rho_1 e^{i \varphi_1}  \\
w^1 &=& \sinh \rho_2 e^{i \varphi_2} \nn  \ , \nn
\eeqa
the generators \eqref{eq:stand-sl2} reduce to
\beqa
\label{eq:s-sl2}
J_+ &=& e^{i\big(\varphi_2 -\varphi_1\big)} \Bigg( -\frac{\sinh \rho_2}{\sinh \rho_1} \partial_{\rho_1} - \frac{\cosh \rho_1}{\cosh \rho_2} \partial_{\rho_2} -
i \frac{\sinh \rho_2}{\cosh \rho_1} \partial_{\varphi_1}  + i \frac{\cosh \rho_1}{\sinh \rho_2} \partial_{\varphi_2} \Bigg) \ , \\
J_- &=& e^{i\big(\varphi_1 -\varphi_2\big)} \Bigg( \frac{\sinh \rho_2}{\sinh \rho_1} \partial_{\rho_1} + \frac{\cosh \rho_1}{\cosh \rho_2} \partial_{\rho_2} -
i \frac{\sinh \rho_2}{\cosh \rho_1} \partial_{\varphi_1}  + i \frac{\cosh \rho_1}{\sinh \rho_2} \partial_{\varphi_2} \Bigg) \ , \nn \\ 
J_0&=&-i\Big(\partial_{\varphi_1} - \partial_{\varphi_2}\Big) \nn \ . 
\eeqa
As in Section \ref{sec:su2}, taking the limit
\beqa
\label{eq:limmm}
\rho_1 \to \rho \ \ , \ \ \text{and} \ \  \rho_2 \to \rho \ ,
\eeqa
in \eqref{eq:s-sl2} reproduces \eqref{eq:sl2H22}. Moreover, in this limit
$w^1, w^2 \in \mathbb H_{2,2} \setminus \mathbb S^1$, hence recovering all the results previously obtained in this section. This means that the representations related to the corresponding
real form of ${\cal S}_\mathbb C^3$ are
related to the standard approach through the limit \eqref{eq:limmm}  and the identification \eqref{eq:sl-w} in \eqref{eq:stand-sl2}.

\section{The algebra of  rotations-translations in two dimensions}
In the previous sections we have realised all simple real or complex  three-dimensional Lie
algebras  using the topological space underlying the corresponding Lie group. 
There is  one more three-dimensional Lie algebra (with a  semi-direct sum structure) which
has interesting properties,  namely the algebra of  translations-rotations in two dimensions. We denote  by
$E_2$ the corresponding group and $\mathfrak{e_2}$ its  Lie algebra.
This group can be obtained,  among other possibilities, by an In\"on\"u-Wigner contraction of $SO(3)$ \cite{IW,GIL}. Indeed, if
we define
\beqa
J= R_0 \ , P_\pm = \e R_\pm \ , \nn
\eeqa
and we take the limit when $\e$ goes to zero, then \eqref{eq:rot} reduces to
\beqa
\label{eq:e2alg}
\big[J,P_\pm\big] = \pm P_\pm\ , \ \ \big[P_+,P_-\big]= 0 \ .
\eeqa
This means that $J$ is the generator of rotations and $P_\pm$ of translations.

 \noindent Since $\pi_1(E_2) = \mathbb Z$ and 
 $E_2$ is non-compact, it shares some properties  with the group
$SL(2,\mathbb R)$: its unitary representations are infinite dimensional and the eigenvalues
of $J$ can be equal to any real number. More  precisely, unitary representations are parameterised by two numbers
$p \in \mathbb R, -1/2 < s \le 1/2$ and are  given by
\beqa
J \big|p,s,n\big> &=& (s+n) |p,s,n\big> \ , \nn \\
P_+ \big|p,s,n\big> &=& i p |p,s,n+1\big> \ , \nn \\
P_- \big|p,s,n\big> &=& -i p |p,s,n-1\big> \ . \nn
\eeqa
The representations are unbounded from below and above since $n \in \mathbb Z$.

\subsection{Representations of $\mathfrak{e}_2$ by contraction}

\noindent Contractions of Lie algebra representations have been
studied from a variety of points of view \cite{TUN}. In contrast to the
contraction of structure tensors, limiting processes for
representations are not straightforward, as they implicitly
involve topological properties of the corresponding Lie groups
\cite{WE}. In this paragraph, unitary representations
of $\mathfrak{e}_{2}$ are reviewed from the perspective of contractions of $\mathfrak{sl}(2,\mathbb
R)$-representations, via the realisations considered before.

\noindent Considering the $\mathfrak{sl}(2,\mathbb R)$ algebra
given by \eqref{eq:sl2H22}, and introducing  $r = \cosh \rho$ we
get  the differential operators

\beqa
iJ_+ &=& \frac{i}2 e^{i(\p_+ - \p_-)}\Big(-i\frac{\sqrt{r^2-1}}{r} \partial_{\p_+} -\sqrt{r^2-1} \partial_r +i
 \frac{r}{\sqrt{r^2-1}} \partial_{\p_-}\Big) \ ,\nn \\
iJ_- &=& \frac{i}2 e^{i(\p_- - \p_+)}\Big(-i\frac{\sqrt{r^2-1}}{r} \partial_{\p_+} +\sqrt{r^2-1} \partial_r +i
 \frac{r}{\sqrt{r^2-1}} \partial_{\p_-}\Big) \ ,\nn \\
J_0 &=& -\frac{i}2\Big(\partial_{\p_+} -\partial_{\p_-}\Big) \ . \nn
\eeqa
Taking the limit  for $r\to + \infty$ we obtain  the realisation
\beqa
\label{eq:lim}
iJ^\infty_+ &=& \frac{i}2 e^{i(\p_+ - \p_-)}\Big(-i \partial_{\p_+} -r \partial_r +i
  \partial_{\p_-}\Big) \ ,\nn \\
iJ^\infty_- &=& \frac{i}2 e^{i(\p_- - \p_+)}\Big(-i \partial_{\p_+} +r \partial_r +i
  \partial_{\p_-}\Big) \ , \\
J^\infty_0 &=& -\frac{i}2\Big(\partial_{\p_+}
-\partial_{\p_-}\Big) \ . \nn 
\eeqa 
 It   satisfies the
$\mathfrak{sl}(2,\mathbb R)$-commutation relations. In the same
limit $\zeta_\pm$ reduces to 
\beqa 
\label{eq:cone-param}
r_+=\zeta^\infty_+= r e^{i
\p_+} \ , \ \ r_-=\zeta_-^\infty = r e^{i\p_-} \ , \ \  r \in \mathbb R_+\ , \p_\pm \in [0,2\pi[ \ ,
\eeqa 
which
belongs to the spinor representation (pay attention
 to the fact that, due to the $i-$factor in $iJ_\pm^\infty$, there is some overall $i$ factor
with respect to Section \ref{sec:repsl2}).

\noindent Now we forget that we have taken the limit $r\to \infty$
and we define the contraction to be 
\beqa
\label{eq:e2}
P_+&=& -\frac{i}2 e^{i(\p_+ - \p_-)} r \partial_r \ ,\nn \\
P_-&=& \frac{i}2 e^{i(\p_- - \p_+)} r \partial_r \ ,\\
J&=&  -\frac{i}2 \Big(\partial_{\p_+} - \partial_{\p_-} \Big) \ .   \nn
\eeqa
One can see easily that these generators generate
the $\mathfrak{e}_2$ algebra and satisfy \eqref{eq:e2alg}.
The parameter space of $E_2$ allows a parameterisation of  the cone ${\cal C}_{2,2}$ 
 since
$[0,2\pi[\times \mathbb R^2 \sim [0,2 \pi[^2 \times \mathbb R_+$.   Here 
$[0,2\pi[\times \mathbb R^2 $
corresponds
respectively to the angle of rotation and the space translation,  whereas  $[0,2\pi[^2\times \mathbb R_+$
allows to obtain a parameterisation of the cone. Indeed \eqref{eq:cone-param}  leads to the equality
\beqa
\Big|r_+\Big|^2 - \Big|r_-\Big|^2 = 0 \ . \nn
\eeqa
Denote ${\cal I}^c_{2,2}= [0,2\pi[\times [0,2\pi[ \times \mathbb R_+^*$. Now, if we remove the point $r=0$
from the cone,
we have a bijection from ${\cal I}^c_{2,2}$ onto ${\cal C}_{2,2}\setminus\big\{0\big\}$
such that the direct application is continuous and the reciprocal is not.
Furthermore,
 since the manifold ${\cal C}_{2,2}\setminus\big\{0\big\}$ is clearly homeomorphic to $\mathbb R_+^*\times \mathbb S^1 \times
\mathbb S^1$, the first homotopy group reduces to $\mathbb Z \times \mathbb Z$. This means that one may
consider covering spaces for ${\cal C}_{2,2}\setminus\big\{0\big\}$ is a similar manner as we have considered
covering spaces for $\mathbb H_{2,2}\setminus \mathbb S^1$.

Finally, one can check that the cone is stable under the action of the generators of ${\mathfrak e}_2$ 
\eqref{eq:e2}, and that $r_\pm$ belongs to the parameter space
of $E_2$ and thus parameterises ${\cal C}_{2,2}\setminus\{0\}$. Of course $(r_+,r_-)$ is not a representation of 
$\mathfrak{e}_2$, but interestingly all unitary representations of the Euclidean Lie algebra 
 in two dimensions can be obtained in a simple way
with $r_+$ and $r_-$. 
If we set for  $-\frac12< s \le \frac12,  p \in \mathbb R$
\beqa
{\cal D}_{s,p} = \Bigg\{
\Lambda_{p,s,n} (r,\p_+,\p_-)=  \frac1{\sqrt{2\pi}}| r_+ r_-|^{p-s-n} r_+^{2s+n} \bar r_-^n =\frac1{\sqrt{2\pi}}r^{2p} e^{i ( 2s+n) \p_+ - i n \p_-}  \ , \ \ n \in \mathbb Z\Bigg\}  \ ,
\ \  
\eeqa
we have the action
\beqa
P_+ \Lambda_{p,s,n} (r,\p_+,\p_-) &=&- ip \Lambda_{p,s,n+1} (r,\p_+,\p_-)  \ , \nn \\
P_- \Lambda_{p,s,n}  (r,\p_+,\p_-)&=& ip \Lambda_{p,s,n-1}  (r,\p_+,\p_-) \ , \nn \\
J \Lambda_{p,s,n} (r,\p_+,\p_-) &=& (s+n) \Lambda_{p,s,n} (r,\p_+,\p_-)  \ . \nn
\eeqa
The functions  $\Lambda_{p,s,n}$ are defined on ${\cal C}_{2,2} \setminus\{0\}
$ when $s=0,1/2$ or eventually on 
one of its coverings if $2s$ 
is not an integer number. More precisely, if $2s =t/q$, the functions are defined on the
$(q,1)-$ sheeted covering $\widetilde{\Big({\cal  C}_{2,2}\setminus\big\{0\big\}\Big)}^{(q,1)}$ 
(with the notations of the covering of $\mathbb H_{2,2}\setminus \mathbb S^1$) parameterised by  
$r\in \mathbb R_+^{ *},0\le \p_+ < 2q\pi, 0 \le \p_- < 2\pi$, 
while for  $s$ a real  non-rational number, the functions are defined on the 
$(\infty,1)-$ sheeted covering $\widetilde{\Big({\cal  C}_{2,2} \setminus\big\{0\big\}\Big)}^{(\infty,1)}$ 
 parameterised by $r\in \mathbb R^{*}_+,\p_+ \in \mathbb R, 0 \le \p_- < 2\pi$. 
 We observe that isomorphic representations could be obtained with 
$| r_+ r_-|^{2p-2s-2n} r_+^{n} \bar r_-^{2s+n}$, where the role of $\p_+$ and $\p_-$ is permuted. We shall however not consider 
this possibility further in detail.

 \noindent Now, with  the change of variables $r=\cosh \rho$, the Laplacian \eqref{eq:lapsl2} becomes
\beqa
\Delta = -\frac1{r} \partial_r(r (r^2-1) \partial_r) + \frac1{r^2}\partial_{\p_+} - \frac1{r^2-1}\partial_{\p_-}
 \ , \nn
\eeqa  and it reduces in the limit $r \to \infty$ to

\beqa \Delta_c =-\frac1{r}\partial_r(r^3 \partial_r) \ . \nn \eeqa
It is straightforward to observe  that  the functions
$\Lambda_{p,s,n}$ are eigenfunctions of $\Delta_c$ \beqa \Delta_c
\Lambda_{p,s,n}(r,\p_+,\p_-)=-(2+2p) 2p\Lambda_{p,s,n}(r,\p_+,\p_-)
\ . \nn
\eeqa

\noindent Note that for $\mathfrak{e}_2$, the Casimir operator is
given by 
\beqa 
Q=\frac12(P_+ P_- + P_- P_+)=\frac14 r \partial(r
\partial_r) \ , \nn \eeqa and thus \beqa Q
\Lambda_{p,s,n}(r,\p_+,\p_-) =  p^2
\Lambda_{p,s,n}(r,\p_+,\p_-) \ . \nn 
\eeqa 
 As it can somehow be expected basing on the non-simplicity of the Euclidean Lie algebra, here 
$\Delta_c$ is not related to the Casimir operator of  $\mathfrak{e}_2$. This result can also be understood {\it a posteriori} because  
the cone ${\cal C}_{2,2}$ is a singular limit of ${\mathbb  H}_{2,2}$. 

\medskip
 To define an invariant scalar product  in this case is more involved, as the eigenvalues of the Casimir operator $Q$
are continuous  hence  the eigenfunctions  $\Lambda_{p,s,n}$ cannot be normalised. However, as we now show using
an appropriate change of parameterisation, we can still define an invariant scalar product.
The first step in the construction is to define the Haar measure for $E_2$ corresponding
to our parameterisation. Using   the fact that 
the cone  is paramaterised by the variables $z_\pm$  together with \eqref{eq:e2}, we can show that under a translation of vector $(a \cos\theta,a\sin \theta)$
we have
\beqa
\Big(r,\varphi_+,\varphi_-\Big) \to
\Big(  (ar)  \cos(-\theta +\varphi_+-\varphi_- - \frac{\pi}2), 
\varphi_+,\varphi_-\Big) \ , \nn
\eeqa
and therefore
\beqa
\text{d} \mu(E_2) = \frac1{r} \text{d} r \text{d} \varphi_+ \text{d} \varphi_- \ , \nn
\eeqa
is an $E_2-$invariant measure.
Now performing a change of variables $r =e^\phi$ with $\Phi \in \mathbb R$, we obtain
\beqa
P_+ &=& -\frac{i}2 e^{i(\varphi_+ -\varphi_-)}\partial_\Phi \ , \nn\\
P_- &=& \frac{i}2 e^{i(\varphi_- -\varphi_+)}\partial_\Phi \ , \nn\\
J&=&-\frac{i}2(\partial_{\p_+} -\partial_{\p_-}) \ , \nn
\eeqa
$\Lambda_{p,s,n}(\Phi,\p_+,\p_-) = e^{2p \Phi} e^{i(n+2s)\p_+ - in\p_-}$ and $\text{d} \mu(E_2) = \text{d} \Phi 
\text{d} \p_+ \text{d} \p_-$.
 This is however still not sufficient  to define an appropriate scalar product.
To proceed further we observe that the Lie algebra $\mathfrak{e}_2$  can also be
obtained by an In\"on\"u-Wigner contraction  of either $\mathfrak{sl}(2,\mathbb R)$ or  $\mathfrak{su}(2)$.
Note also that both algebras are real forms of the  complex $\mathfrak{sl}(2,\mathbb C)$  Lie algebra.
Formally, to go from one real form to the other we just have to perform the following substitution
\beqa
\label{eq:sub}
J_0^{\mathfrak{sl}(2,\mathbb R)} &\to& J_0^{\mathfrak{su}(2)} = J_0^{\mathfrak{sl}(2,\mathbb R)} \ , \nn \\
J_\pm^{\mathfrak{sl}(2,\mathbb R)} &\to& J_\pm^{\mathfrak{su}(2)} = -i J_\pm^{\mathfrak{sl}(2,\mathbb R)} \ ,\ 
\eeqa
 or some other equivalent transformation as those used in the theory of Angular Momentum \cite{LOU}.
In our procedure to construct $\mathfrak{e}_2$ and its unitary representations we  have considered the contraction
of $\mathfrak{sl}(2,\mathbb R)$. Going from $\mathfrak{sl}(2,\mathbb R)$ to $\mathfrak{su}(2)$
through the substitution \eqref{eq:sub} can just be 
realised  by $\Phi \to \Psi = -i \Phi$ with $\Psi \in \mathbb R$.  
With this substitution, the variables $r_\pm$ are now living on the manifold 
$\mathbb S^1 \times \mathbb S^1 \times \mathbb S^1$.  Thus, 
in the transition from $\Phi\in \mathbb R$ to $\Psi \in \mathbb S^1$  
 we have substituted 
 $\mathbb R \times \mathbb S^1 \times \mathbb S^1$ by $\mathbb S^1 \times \mathbb S^1 \times \mathbb S^1$. 
Meaning that on a formal ground $\mathbb S^1 \times \mathbb S^1 \times \mathbb S^1$ can be seen as
the one point compactification of $\mathbb R \times \mathbb S^1 \times \mathbb S^1$ since
the first circle is the one-point compactification of $\mathbb R$. Now since the latter manifold is compact
a scalar product can be defined.
The parameterisation
$(\Psi,\p_+,\p_-) \in [0,2\pi[\times [0,2\pi[\times [0,2\pi[$ uniquely defines a point
on $\mathbb S^1 \times \mathbb S^1 \times \mathbb S^1$,  
the first homotopy group of which is isomorphic to $\mathbb Z\times\mathbb Z\times
\mathbb Z$. This leads to the new realisation of $\mathfrak{e_2}$
\beqa
P_+ &=& -\frac{1}2 e^{i(\varphi_+ -\varphi_-)}\partial_\Psi \ , \nn\\
P_- &=& \frac{1}2 e^{i(\varphi_- -\varphi_+)}\partial_\Psi \ , \nn\\
J&=&-\frac{i}2(\partial_{\p_+} -\partial_{\p_-}) \ . \nn
\eeqa
and of
the representations ${\cal D}_{s,p}$
\beqa
\Lambda_{p,s,n}(\Psi,\p_+,\p_-) = e^{i 2p \Psi} e^{i(n+s)\p_+ -in  \p_-} \ . \nn
\eeqa
The representations of $E_2$ are defined on some covering of $\mathbb S^1\times\mathbb S^1\times
\mathbb S^1$ like before, namely $ \widetilde{ \mathbb S^1}^\infty\times\widetilde{\mathbb S^1}^p\times\mathbb S^1$
or $ \widetilde{ \mathbb S^1}^\infty\times\widetilde{\mathbb S^1}^\infty\times\mathbb S^1$, where
 $\widetilde{ \mathbb S^1}^p$ (resp.  $\widetilde{ \mathbb S^1}^\infty$) is the $p-$sheeted (universal) covering
of the circle $ \mathbb S^1$. These spaces are defined by their parameterisation.
If $s=0,1/2$  the functions  $\Lambda_{p,r/q,n}$ are  parameterised by $ \Psi \in \mathbb R, \p_+ \in
[0,2\pi[, \p_-\in[0,2\pi[$, if
$2s = r/q$ by $ \Psi \in \mathbb R, \p_+ \in
[0,2q\pi[, \p_-\in[0,2\pi[$ and if $s \in \mathbb R \setminus \mathbb Q$ by 
 $ \Psi \in \mathbb R, \p_+ \in \mathbb R, \p_-\in[0,2\pi[$. 

\medskip
\noindent 
Now, considering two functions $f$ and $g$ 
 defined on  some covering of $\mathbb S^1\times \mathbb S^1 \times \mathbb S^1$,  we can endow the latter space with the scalar product
\beqa
(f,g)_q =\frac1{q}\frac1{(2 \pi)^2} \int \limits_{-\infty}^{+\infty} \text{d} \Psi \int 
\limits_0^{2 q\pi} \text{d} \p_+ \int \limits_0^{2 \pi} \text{d} \p_-
\bar f(\Psi,\p_+,\p_-) g(\Psi,\p_+,\p_-) \ , \nn
\eeqa
 and
\beqa
(f,g)_\infty =\frac2{(2 \pi)^2} \int \limits_{-\infty}^{+\infty} \text{d} \Psi \int 
\limits_{-\infty}^{+\infty} \text{d} \p_+ \int \limits_{0}^{2 \pi} \text{d} \p_-
\bar f(\Psi,\p_+,\p_-) g(\Psi,\p_+,\p_-) \ , \nn
\eeqa
 respectively. Using the integral representation
of the Dirac $\delta-$distribution we have  the following situation:
for  $s,s'\in \mathbb Q$, {\it i.e.}, $2s=r/q, 2s'=r'/q'$,
and considering  $q''$ the least common multiple of $q,q'$ 
\beqa
(\Lambda_{p',\frac{r'}{2q'},n'}, \Lambda_{p,\frac{r}{2q},n})_{q''} = \delta_{nn'} \delta_{\frac{r}{q} \frac{r'}{q'}}
\delta(p-p') \ . \nn
\eeqa
If $s$ or $s'$ is a real number (but not a rational number)
\beqa
(\Lambda_{p',s',n'}, \Lambda_{p,s,n})_{q''} = \delta_{nn'} \delta(s'-s) 
\delta(p-p') \ . \nn
\eeqa
The functions $\Lambda_{p,s,n}$ are  hence orthonormal, proving that the representations ${\cal D}_{p,s}$
are unitary.

\section{Final remarks}

\noindent 
In this paper the Gel'fand formul\ae \ for the unitary representations of the $SL(2,\mathbb C)$ Lie
group have been  reviewed, emphasising the description of  unitary  representations in terms of harmonic functions
on the spaces  $ \mathbb S^3_\mathbb C \setminus\Big({\cal C}_2 \times {\cal C}_2\Big)\subset SL(2,\mathbb C)$
. Considering appropriate real forms of  ${\mathbb S}_\mathbb C^3\setminus\Big({\cal C}_2 \times {\cal C}_2\Big)$ further enables us to adapt the previous formul\ae \ in such way that they reduce to unitary representations of
the real forms of $SL(2,\mathbb C)$, say $SU(2)$ and $SU(1,1)$. At the same time, the corresponding representations
become harmonic functions  on $\mathbb S^3\setminus\Big(\mathbb S^1\times \mathbb S^1\Big) \subset 
SU(2)$ or
$\mathbb H_{2,2}\setminus \mathbb S^1 \subset SL(2,\mathbb R)$ (or one of
its covering space)  respectively.  In addition, considering a suitable contraction of
$SU(2)$ allows to derive the unitary representations of the Euclidean group $E_2$ on
 ${\cal C}_{2,2}\setminus\{0\}$, or its one-point compactification
$\mathbb S^1 \times \mathbb S^1 \times \mathbb S^1$.  In this frame,  it is
important to realise  that for the Lie groups $SL(2,\mathbb R)$ and $E_2$, being both infinitely connected, representations
on their $p-$sheeted covering can be defined (corresponding to a real spin). Interestingly, defining appropriate
coverings of  either ${\mathbb H}_{2,2} \setminus \mathbb S^1$ 
or ${\cal C}_{2,2} \setminus \{0\}$, the representations of  $p-$sheeted coverings can safely be defined.

\medskip
\noindent 
This approach could be potentially useful for the analysis
of other compact Lie groups like
$SU(N)$, as well as 
an alternative procedure to construct bases of eigenstates for physically relevant chains like $SU(N)\supset \cdots \supset SO(3)$ and the study of
subduced representations in terms of harmonics, in analogy to the case studied in  \cite{GR}.

\noindent Let us finally mention that three-dimensional  simple Lie groups are somehow exceptional, in the sense that all 
their representations can be constructed from the spinor representation(s). Within the construction revisited here,
spinors are living on a suitable manifold related to the Lie groups themselves, and since spinors are harmonic, all unitary
representations will be harmonic  on the corresponding manifold. This property is non longer valid for higher dimensional Lie
groups. For instance, for $SU(3)$, representations  are obtained as harmonic functions on the five-sphere
\cite{Beg},  which no longer coincides with  a manifold related to the Lie group $SU(3)$, 
by considering the three-dimensional representation
and its complex conjugate. Basing on the latter observation, a similar ansatz could certainly be considered for Lie groups of  rank higher
than two and different from $SO(n)$, like $SU(n), n>3$ or $Sp(N), N>2$.  However, it is to be expected that in this case,  the computational difficulties will certainly prevent us from obtaining  all the representations in closed form as harmonic functions on some appropriate manifold.

\subsection*{Acknowledgments}
 The authors express their gratitude to Marcus Slupinski for fruitful discussions and valuable suggestions that improved the manuscript. 
 This work was partially supported by the research project MTM2013-43820-P of the MINECO (Spain).

\bibliographystyle{utphys}
\bibliography{biblio}

\end{document}